\documentclass{article}
\usepackage{graphicx}
\usepackage{amsmath}

\input{tcilatex}

\begin{document}

\author{Sawa Manoff \\
\textit{Bulgarian Academy of Sciences}\\
\textit{Institute for Nuclear Research and Nuclear Energy}\\
\textit{Department of Theoretical Physics}\\
\textit{Blvd. Tzarigradsko Chaussee 72}\\
\textit{1784 Sofia - Bulgaria}}
\title{\textsc{Lagrangian theory for perfect fluids} }
\date{e-mail address: smanov@inrne.bas.bg}
\maketitle

\begin{abstract}
\textit{The theory of perfect fluids is reconsidered from the point of view
of a covariant Lagrangian theory. It has been shown that the Euler-Lagrange
equations for a perfect fluid could be found in spaces with affine
connections and metrics from an unconstrained variational principle by the
use of the method of Lagrangians with covariant derivatives (MLCD) and
additional conditions for reparametrizations of the proper time of the mass
elements (particles) of the perfect fluid. The last conditions are not
related to the variational principle and are not considered as constraints
used in the process of variations. The application of the whole structure of
a Lagrangian theory with an appropriate choice of a Lagrangian invariant as
the pressure of the fluid shows that the Euler-Lagrange equations with their
corresponding energy-momentum tensors lead to Navier-Stokes' equation
identical with the Euler equation for a perfect fluid in a space with one
affine connection and metrics. The Navier-Stokes equations appear as higher
order equations with respect to the Euler-Lagrange equations.}
\end{abstract}

\section{Introduction}

1. Every classical field theory could be considered as a theory of a
continuous media \cite{Manoff-1} $\div $ \cite{Manoff-4a} with its kinematic
characteristics. On the one side, a classical field theory is usually based
on three essential structures:

(a) The Lagrangian density,

(b) The Euler-Lagrangian equations,

(c) The energy-momentum tensors.

The structures in (a) - (c) determine the scheme of a Lagrangian formalism
for describing the properties and the evolution of a dynamic system. The
finding out of the Euler-Lagrange equations and the energy-momentum tensors
is related to a variational principle applied under certain mathematical
conditions for the variables and their variations \cite{Manoff-3}, \cite
{Manoff-4a}.

On the other side, a theory of a fluid is based on

(a) A state equation

(b) A variational principle with a given Lagrangian density depending on the
variables constructing the equation of state (rest mass density, entropy,
enthalpy, temperature, pressure etc.) and the velocity of the points of the
fluid.

2. Since a theory of a fluid is a special case of a theory of continuous
media, it could be worked out by means of the same mathematical tools used
in a classical field theory. Moreover, the Lagrangian invariant $L$ in the
construction of the Lagrangian density $\mathbf{L}:=\sqrt{-d_g}\cdot L$, $%
d_g:=det(g_{ij})<0$, could be interpreted as the pressure $p$ of a physical
system, described by its Lagrangian density $\mathbf{L}$. Here, the tensor $%
g=g_{ij}\cdot dx^i.dx^j$, $dx^i.dx^j:=\frac 12\cdot (dx^i\otimes
dx^j+dx^j\otimes dx^i)$ is the metric tensor field in a differentiable
manifold $M$, $dimM=n$, considered as a model of the space $(n=3)$ or of the
space-time $(n=4)$. The fact that the pressure $p$ could be used as
Lagrangian invariant $L=p$ has been observed by many authors \cite{Schutz-2}
(see the citations there) but used only for very special forms of the
pressure $p$. The application of the assumption $L:=p$ could be extended to
many classical field theories and to theories of different types of fluids 
\cite{Manoff-4a}.

3. Every Lagrangian density depends on a set of field variables. Part of
them (or all of them) are considered as dynamic field variables, i.e. as
variables which variations lead to the Euler-Lagrange equations or to
energy-momentum tensors. If a field variable is not varied it is called a
non-dynamic field variable. It depends on the problem to be solved which
variables are considered as dynamic and which as non-dynamic field
variables. Only in the case of finding out the energy-momentum tensors the
Lie variation \cite{Manoff-3} of all field variables should be taken into
account. If all dynamic field variables are independent to each other, there
are no additional conditions between them. The variables are varied without
any constraints and the variational principle used in that case is called 
\textit{unconstrained variational principle}. If a part of the dynamic field
variables are depending in some way to each other there are relations
(constraints) between them. The variation of the variables have to include
the constraints between them and the variational principle used in this case
is called \textit{constrained variational principle}. In many cases the
constraints have no obvious physical interpretation or lead to additional
problems. There are attempts for finding out conditions which could show us
which field theory could be derived from an unconstrained variational
principles and which cannot \cite{Schutz-2}. On this basis Schutz and Sorkin
came to the conclusion that the theory of a perfect fluid could not be
derived from an unconstrained variational principle. At that, they used the
method of Lagrangians with partial derivatives (MLPD) \cite{Manoff-3}. The
theory of a perfect fluid [related to the Euler equation in hydrodynamics
(1727-1741)] is considered as the oldest of classical field theories \cite
{Schutz-2} and it is worth to be investigated by the use of some of the
modern mathematical tools with respect to its structures and elements.

4. In the present paper we have used the method of Lagrangians with
covariant derivatives (MLCD) instead of the method of Lagrangians with
partial derivatives (MLPD) to reconsider the problem of derivation of the
theory of a perfect fluid from an unconstrained variational principle. It
has been shown that the Euler-Lagrange equations for a perfect fluid could
be found in spaces with affine connections and metrics \cite{Manoff-4} from
an unconstrained variational principle by the use of the method of
Lagrangians with covariant derivatives (MLCD) and additional conditions for
reparametrizations of the proper time of the mass elements (particles) of
the perfect fluid. The last conditions are not related to the variational
principle and are not considered as constraints used in the process of
variations. The application of the whole structure of a Lagrangian theory
with an appropriate choice of a Lagrangian invariant as the pressure of the
fluid shows that the Euler-Lagrange equations with their corresponding
energy-momentum tensors lead to Navier-Stokes' equations containing the
Euler equation for a perfect fluid as a special case in a space with affine
connections and metrics \cite{Manoff-0}, \cite{Manoff-4}. The Navier-Stokes
equation appear as higher order equations with respect to the Euler-Lagrange
equations.

5. The structure of the paper is as follows. In Section 2 a brief recall of
the method of Lagrangians with covariant derivatives (MLCD) is made and
applied to a Lagrangian density depending on a metric, a scalar field, and
on a contravariant non-isotropic (non-null) vector field. In Section 3 the
invariant projections of the energy-momentum tensors are considered for the
corresponding Lagrangian density.. In Section 4 the covariant divergencies
of the energy-momentum tensors and their relations to Navier-Stokes'
equations are given. In Section 5 the invariance of dynamic characteristics
under changing the proper time of the trajectory of a mass element
(particle) in a fluid is considered. In Section 6 the notion of perfect
fluid and its mathematical models are considered. In Section 7 the method of
Lagrangians with covariant derivatives is applied to a special type of a
Lagrangian density related to a perfect fluid in special types of $(%
\overline{L}_n,g)$-spaces and in $(L_n,g)$-spaces and the corresponding
Euler-Lagrange equations and energy-momentum tensors are found. In Section 8
the Navier-Stokes equation is considered in $(L_n,g)$-spaces. It is shown
that this equation is identical with the Euler equation if considered for a
perfect fluid with the given Lagrangian density in a $(L_n,g)$-space. The
final Section 9 comprises some concluding remarks. The most considerations
are given in details (even in full details) for these readers who are not
familiar with the considered problems. If some details or abbreviations are
not defined or introduced in the paper the reader is kindly asked to refer
to \cite{Manoff-4} and \cite{Manoff-4a} where all preliminary informations
for reading this paper are given.

\section{Method of Lagrangians with covariant derivatives (MLCD)}

\subsection{General remarks}

The method of Lagrangians with covariant derivatives (MLCD) is considered in
details in \cite{Manoff-3}, \cite{Manoff-5}, \cite{Manoff-4a}. We will
recall here only the main structures and properties of this method.

The \textit{method of Lagrangians with covariant derivatives (MLCD)} is a
Lagrangian formalism for tensor fields based on:

(a) A Lagrangian density $\mathbf{L}$ of type

\begin{equation}
\mathbf{L}=\sqrt{-d_g}.L(g_{ij},g_{ij;k},g_{ij;k;l},V_{\,\,\,\,\,\,B}^A,V_{%
\,\,\,\,\,\,\,\,B;i}^A,V_{\,\,\,\,\,\,B;i;j}^A)\text{ ,}  \label{1.1}
\end{equation}
where $g_{ij}$ are the components of the covariant metric tensor field $g$, $%
_{;k}$ denotes covariant derivative with respect to the co-ordinate $x^k$
(or with respect to the basic co-ordinate vector field $\partial _k$), $%
V^A\,_B$ are components of tensor fields $V\in \otimes ^k\,_l(M)$ with
finite rank, $A$ and $B$ are co-operative indices.

(b) The action $S$ of a Lagrangian system described by means of the
Lagrangian density $\mathbf{L}$%
\begin{equation}
S=\dint\limits_{V_n}\mathbf{L}.d^{(n)}x=\dint\limits_{V_n}L.d\omega \text{ ,}
\label{1.2}
\end{equation}
where $V_n$ is a volume in the manifold $M$ with $\dim M=n$; $d\omega =\sqrt{%
-d_g}.d^{(n)}x$ is the invariant volume element.

(c) The functional variation $\delta S$ of the action $S$ with the condition
for the existence of an extremum 
\begin{equation}
\delta S=\delta \dint\limits_{V_n}\mathbf{L}.d^{(n)}x=\dint\limits_{V_n}%
\delta \mathbf{L}.d^{(n)}x=0\text{ .}  \label{1.3}
\end{equation}

(d) The functional variation of the Lagrangian density $\mathbf{L}$ in the
form 
\begin{equation*}
\delta \mathbf{L}=\frac{\partial \mathbf{L}}{\partial g_{ij}}.\delta g_{ij}+%
\frac{\partial \mathbf{L}}{\partial g_{ij;k}}.\delta (g_{ij;k})+\frac{%
\partial \mathbf{L}}{\partial g_{ij;k;l}}.\delta (g_{ij;k;l})+ 
\end{equation*}
\begin{equation}
+\frac{\partial \mathbf{L}}{\partial V^A\,_B}.\delta V^A\,_B+\frac{\partial 
\mathbf{L}}{\partial V^A\,_{B;i}}.\delta (V^A\,_{B;i})+\frac{\partial 
\mathbf{L}}{\partial V^A\,_{B;i;j}}\delta (V^A\,_{B;i;j})\text{ .}
\label{1.4}
\end{equation}

(e) The variational operator $\delta $ obeying the commutation relations
leading to commutation of $\delta $ with the covariant derivatives: 
\begin{equation}
\begin{array}{c}
\delta (g_{ij;k})=(\delta g_{ij})_{;k}\text{ ,\thinspace \thinspace
\thinspace \thinspace \thinspace }\delta (g_{ij;k;l})=(\delta g_{ij})_{;k;l}%
\text{ ,} \\ 
\delta (V^A\,_{B;i})=(\delta V^A\,_B)_{;i}\text{ , \thinspace \thinspace
\thinspace }\delta (V^A\,_{B;i;j})=(\delta V^A\,_B)_{;i;j}\text{ .}
\end{array}
\label{1.5}
\end{equation}

(f) The Lie variation $\pounds _\xi S$ of the action $S$%
\begin{equation}
\pounds _\xi S=\pounds _\xi \dint\limits_{V_n}\mathbf{L}.d^{(n)}x=\dint%
\limits_{V_n}\overline{\pounds }_\xi \mathbf{L}.d^{(n)}x\text{ \thinspace .}
\label{1.6}
\end{equation}

(g) The Lie variation of the Lagrangian density $\mathbf{L}$ in the form 
\begin{equation*}
\overline{\pounds }_\xi \mathbf{L}\equiv \frac{\partial \mathbf{L}}{\partial
g_{ij}}.\pounds _\xi g_{ij}+\frac{\partial \mathbf{L}}{\partial g_{ij;k}}%
.\pounds _\xi (g_{ij;k})+\frac{\partial \mathbf{L}}{\partial g_{ij;k;l}}%
.\pounds _\xi (g_{ij;k;l})+ 
\end{equation*}
\begin{equation}
+\frac{\partial \mathbf{L}}{\partial V^A\,_B}.\pounds _\xi V^A\,_B+\frac{%
\partial \mathbf{L}}{\partial V^A\,_{B;i}}.\pounds _\xi (V^A\,_{B;i})+\frac{%
\partial \mathbf{L}}{\partial V^A\,_{B;i;j}}.\pounds _\xi (V^A\,_{B;i;j})%
\text{ .}  \label{1.7}
\end{equation}

\textit{Remark}. In the MLCD [because of the commutation relations] \textit{%
the affine connections }$\Gamma $\textit{\ and }$P$,\textit{\ and their
corresponding curvature tensors are considered as non-dynamic field
variables }$(\delta \Gamma _{jk}^i=0$, $\delta P_{jk}^i=0$, $\delta
R^i\,_{jkl}=0$, $\delta P^i\,_{jkl}=0)$\textit{.} Therefore, in the MLCD
variations of the components and their covariant derivatives of the
covariant metric tensor $g$ and the non-metric tensor fields $V$ are
considered for given (fixed) affine connections and for fixed and determined
by them types of transports of the tensor fields. Of course, the question
arises how the affine connections can be found if not by means of a
Lagrangian formalism. The first simple answer is: the affine connections (or
the equations for them as functions of the co-ordinates in $M$) can be found
on the grounds of the method of Lagrangians with partial derivatives (MLPD)
and then the components of the tensor fields (as functions of the
co-ordinates in $M$) can be determined by means of MLCD. This answer could
induce an other question: why two methods have to be applied when one is
enough for finding out all equations for all dynamic field variables. There
are at least two possible answers to this question: 1. The MLCD ensure the
finding out equations for tensor fields (as dynamic field variables). These
equations are (a) covariant (tensorial) equations and (b) form-invariant
(gauge invariant) equations with respect to the affine connections. The
affine connections, related to the type of the frame of reference \cite
{Manoff-6} could be determined on the grounds of additional conditions and
not exactly by means of a variational principle. 2. The MLPD can ensure the
consideration of the affine connections as dynamic field variables and the
finding out their field equations. It cannot give direct answer for the type
(tensorial or non-tensorial) of the equations obtained for the tensor field
variables and their corresponding energy-momentum quantities. The tensorial
character of quantities and relations concerning the tensor fields variables
has to be proved (which, in general, could be a matter of some difficulty) 
\cite{Lovelock}.

\subsection{Method of Lagrangians with covariant derivatives applied to a
scalar and to a vector field}

\subsubsection{Lagrangian density}

1. The mathematical description of a state of a fluid is usually made by
means of the velocity vector field $u$ of the moving mass elements
(particles) of the fluid and other two functions (for instance, the pressure 
$p$ and the rest mass density $\rho $) \cite{Landau}. If $u$, $\rho $, and $%
p $ are given then the state of a fluid is determined. This fact gives rise
to the idea a Lagrangian invariant $L=p$ depending on the velocity vector
field $u$ and on the rest mass density $\rho $ to be used in the MLCD for
finding out the Euler-Lagrange equations and their corresponding
energy-momentum tensors in a space with affine connections and metrics [$(%
\overline{L}_n,g)$-space] (considered as a model of space or space-time).

2. Let a Lagrangian density of the type 
\begin{equation}
\mathbf{L}:=\sqrt{-d_g}\cdot p(g_{ij}\text{, }\rho \text{, }\rho _{;i}\equiv
\rho _{,i}\text{, }u^i)\text{ }  \label{1.8}
\end{equation}
be given, where $d_g:=det(g_{ij})<0$, $g_{ij}=g_{ji}(x^k)$ are the
components of a covariant metric $g$, $\rho $ is an invariant function on $M$%
, $\rho \in C^r(M)$, $r\geq 2$, $\rho \in \otimes ^0\,_0(M)$, $\rho
_{;i}\equiv \rho _{,i}:=\partial \rho /\partial x^i$, $u^i$ are the
components of a non-null (non-isotropic) vector field $u\in T(M)$ in a
co-ordinate basis $\{\partial _i\}\subset T(M)$, $u=u^i\cdot \partial _i$, $%
g(u,u):=e:\neq 0$, $i,j=1,...,n$, $dimM=n$. The Lagrangian invariant $L:=p$
could be interpreted as the pressure $p$ of the fluid. The invariant
function $\rho $ could be interpreted as the rest mass density of the fluid
corresponding to one of its energy-momentum tensors \cite{Manoff-3}. The
contravariant non-isotropic (non-null) vector field $u$ and its components $%
u^i$ could be interpreted as the velocity vector field of the mass elements
(particles) of the fluid.

By the use of the MLCD we can find the Euler-Lagrange equations and the
energy-momentum tensors corresponding to $\mathbf{L}$. The Lagrangian
density $\mathbf{L}$ is a degenerated Lagrangian density with respect to $u^i
$ and $g_{ij}$, i.e. $\mathbf{L}$ contains no derivatives of $u$ and $g_{ij}$
with respect to the co-ordinates $x^i$ or to the basic vector fields $%
e_i\equiv \partial _i$.

\subsubsection{Euler-Lagrange's equations}

\textit{(a) Euler-Lagrange's equations for the rest mass density }$\rho $

The ELE for the rest mass density $\rho $ could be found in the form \cite
{Manoff-3}, \cite{Manoff-4a} 
\begin{equation}
\frac{\delta _\rho p}{\delta \rho }+P=0\,\,\,\,\,\,\text{,}  \label{1.9}
\end{equation}
where 
\begin{eqnarray}
\frac{\delta _\rho p}{\delta \rho } &=&\frac{\partial p}{\partial \rho }-(%
\frac{\partial p}{\partial \rho _{;i}})_{;i}\text{ \thinspace \thinspace
\thinspace ,}  \label{1.10} \\
P &=&q_i\cdot \frac{\partial p}{\partial \rho _{;i}}\text{ \thinspace
\thinspace \thinspace \thinspace \thinspace ,\thinspace \thinspace
\thinspace \thinspace \thinspace \thinspace \thinspace \thinspace \thinspace
\thinspace }q_i=T_{ki}\,^k-\frac 12\cdot g^{\overline{k}\overline{l}}\cdot
g_{kl;i}+g_k^l\cdot g_{l;i}^k\text{ \thinspace \thinspace .}  \label{1.10a}
\end{eqnarray}

Therefore, the ELE for $\rho $ could be written as 
\begin{equation}
\frac{\partial p}{\partial \rho }-(\frac{\partial p}{\partial \rho _{;i}}%
)_{;i}+q_i\cdot \frac{\partial p}{\partial \rho _{;i}}=0\text{ \thinspace
\thinspace \thinspace .}  \label{1.11}
\end{equation}

\textit{(b) Euler-Lagrange's equations for the vector field }$u$

The Lagrangian density $\mathbf{L}$ is degenerated with respect to the
vector field $u$, i.e. the ELEs could be written in the form \cite{Manoff-4}%
, \cite{Manoff-4a} 
\begin{equation}
\frac{\partial p}{\partial u^i}=0\,\,\,\,\,\,\text{.}  \label{1.12}
\end{equation}

\textit{(c) Euler-Lagrange's equations for the metric tensor }$g$

The Lagrangian density $\mathbf{L}$ is degenerated with respect to the
components $g_{ij}$ of the tensor field $g$. Therefore, 
\begin{equation*}
\frac{\delta _gp}{\delta g_{ij}}=\frac{\partial p}{\partial g_{ij}}\text{ ,} 
\end{equation*}
and we have the ELEs in the form 
\begin{equation}
\frac{\partial p}{\partial g_{ij}}+\frac 12\cdot p\cdot g^{\overline{i}%
\overline{j}}=0\text{ \thinspace \thinspace \thinspace .}  \label{1.13}
\end{equation}

If the ELEs for $g_{ij}$ are fulfilled then the pressure $p$ should obey the
equation 
\begin{equation}
\frac{\partial p}{\partial g_{ij}}\cdot g_{ij}+\frac n2\cdot
p=0\,\,\,\,\,\,\,\,:\,\,\,\,\,\,\,\,\,\,p=-\frac 2n\cdot \frac{\partial p}{%
\partial g_{ij}}\cdot g_{ij}\text{ \thinspace \thinspace .}  \label{1.14}
\end{equation}

The last condition means that $p$ should be a homogeneous function of $%
g_{ij} $ of degree $m=-(2/n)$. If the pressure $p$ does not have this
property then the metric tensor $g$ could not be considered as a dynamic
variable, i.e. it should not be varied.

\subsubsection{Energy-momentum tensors}

In accordance to the general scheme of the MLCD we could find the explicit
form of the corresponding to the pressure $p$ energy-momentum tensors.

\textit{(a) Generalized canonical energy-momentum tensor}

\begin{equation}
\overline{\theta }_i\,^j=\frac{\partial p}{\partial \rho _{;j}}\cdot \rho
_{;i}-p\cdot g_i^j\,\,\,\,\,\text{.}  \label{1.15}
\end{equation}

\textit{(b) Symmetric energy-momentum tensor of Belinfante} 
\begin{equation}
_sT_i\,^j=-p\cdot g_i^j\text{ \thinspace \thinspace \thinspace \thinspace .}
\label{1.16}
\end{equation}

\textit{(c) Variational energy-momentum tensor of Euler-Lagrange}

\begin{eqnarray*}
\overline{Q}_i\,^j &=&\frac{\delta _up}{\delta u^k}\cdot g_i^k\cdot
g_l^j\cdot u^l-\frac{\delta _gp}{\delta g_{kl}}\cdot g_k^{\underline{j}%
}\cdot g_{\overline{i}}^m\cdot g_{ml}-\frac{\delta _gp}{\delta g_{kl}}\cdot
g_l^{\underline{j}}\cdot g_{\overline{i}}^m\cdot g_{km}= \\
&=&\frac{\partial p}{\partial u^i}\cdot u^j-\frac{\partial p}{\partial g_{kl}%
}\cdot g_k^{\underline{j}}\cdot g_{\overline{i}}^m\cdot g_{ml}-\frac{%
\partial p}{\partial g_{kl}}\cdot g_l^{\underline{j}}\cdot g_{\overline{i}%
}^m\cdot g_{km}\text{ \thinspace \thinspace .}
\end{eqnarray*}

Because of $g_{\overline{i}k}=f^m\,_i\cdot g_{mk}=f^m\,_i\cdot g_{km}=g_{k%
\overline{i}}$, \thinspace $g_{kl}=g_{lk}$, $g_{\overline{i}}^m\cdot
g_{ml}=f^n\,_i\cdot g_n^m\cdot g_{ml}=f^n\,_i\cdot g_{nl}=g_{\overline{i}l}$
we have 
\begin{equation}
\overline{Q}_i\,^j=\frac{\partial p}{\partial u^i}\cdot u^j-2.\frac{\partial
p}{\partial g_{kl}}\cdot g_{\overline{i}k}\cdot g_l^{\underline{j}}\text{
\thinspace \thinspace \thinspace \thinspace .}  \label{1.17}
\end{equation}

\section{Invariant projections of the energy-momentum tensors}

Since the vector field $u$ could be interpreted as the velocity of the mass
elements (particles) of a fluid, we can project the energy-momentum tensors $%
\overline{\theta }_i\,^j$, $_sT_i\,^j$, and $\overline{Q}_i\,^j$ by the use
of the vector field $u$ and its corresponding covariant and contravariant
projective metrics $h_u$ and $h^u$%
\begin{eqnarray}
h_u &=&g-\frac 1e\cdot g(u)\otimes g(u)\text{ \thinspace \thinspace
\thinspace ,\thinspace \thinspace \thinspace \thinspace \thinspace
\thinspace \thinspace \thinspace \thinspace \thinspace }h^u=\overline{g}%
-\frac 1e\cdot u\otimes u\text{ \thinspace \thinspace \thinspace ,}
\label{2.1} \\
e &=&g(u,u):\neq 0\text{ \thinspace \thinspace \thinspace ,\thinspace
\thinspace \thinspace \thinspace \thinspace \thinspace }\overline{g}%
=g^{ij}\cdot \partial _i.\partial _j\text{ \thinspace \thinspace \thinspace ,%
}  \notag \\
\text{\thinspace \thinspace \thinspace \thinspace \thinspace }\partial
_i.\partial _j &=&\frac 12\cdot (\partial _i\otimes \partial _j+\partial
_j\otimes \partial _i)\text{ ,}  \notag \\
g(u) &=&g_{i\overline{j}}\cdot u^j\cdot dx^i=g_{ij}\cdot u^{\overline{j}%
}\cdot dx^i\text{ \thinspace \thinspace \thinspace .}  \notag
\end{eqnarray}

By the use of $u$, $h_u$, and $h^u$ we can find the different invariant
projections of $\overline{\theta }_i\,^j\,$,$\,\,_sT_i\,^j$, and $\overline{Q%
}_i\,^j$ which have physical interpretation.

\subsection{Invariant projections of the generalized canonical
energy-momentum tensor $\overline{\protect\theta }$}

1. Rest mass density $\rho _\theta $. 
\begin{eqnarray}
\rho _\theta &=&\frac 1{e^2}\cdot \overline{\theta }_k\,^i\cdot u_{\overline{%
i}}\cdot u^{\overline{k}}=  \notag \\
&=&\frac 1{e^2}\cdot u_{\overline{i}}\cdot u^{\overline{k}}\cdot (\frac{%
\partial p}{\partial \rho _{;i}}\cdot \rho _{;k}-p\cdot g_k^i)=  \notag \\
&=&\frac 1{e^2}\cdot (\,\frac{\partial p}{\partial \rho _{;i}}\cdot u_{%
\overline{i}}\cdot u^{\overline{k}}\cdot \rho _{;k}-p\cdot u_{\overline{k}%
}\cdot u^{\overline{k}})\text{ .}  \label{2.2}
\end{eqnarray}

2. Conductive momentum density $^\theta \overline{\pi }$. 
\begin{equation}
^\theta \overline{\pi }\,^i:=\frac 1e\cdot \,_k\overline{\theta }_l\,^k\cdot
u_{\overline{k}}\cdot h^{\overline{l}i}=\frac 1e\cdot \frac{\partial p}{%
\partial \rho _{;k}}\cdot \rho _{;l}\cdot u_{\overline{k}}\cdot h^{\overline{%
l}i}\text{ \thinspace \thinspace \thinspace .}  \label{2.3}
\end{equation}

3. Conductive energy flux density $e\cdot \,^\theta \overline{s}$. 
\begin{equation}
^\theta \overline{s}^i:=\frac 1e\cdot h^{ij}\cdot g_{\overline{j}\overline{k}%
}\cdot \,_k\overline{\theta }_l\,^k\cdot u^{\overline{l}}=\frac 1e\cdot
h^{ij}\cdot g_{\overline{j}\overline{k}}\cdot \frac{\partial p}{\partial
\rho _{;k}}\cdot \rho _{;l}\cdot u^{\overline{l}}\,\,\,\,\,\text{.}
\label{2.4}
\end{equation}
\begin{equation}
e\cdot \,^\theta \overline{s}^i=h^{ij}\cdot g_{\overline{j}\overline{k}%
}\cdot \frac{\partial p}{\partial \rho _{;k}}\cdot \rho _{;l}\cdot u^{%
\overline{l}}\,\,\,\,\,\text{.}  \label{2.5}
\end{equation}

4. Stress tensor density $^\theta \overline{S}$. 
\begin{equation}
^\theta \overline{S}^{ij}:=h^{ik}\cdot g_{\overline{k}\overline{l}}\cdot \,_k%
\overline{\theta }_m\,^l\cdot h^{\overline{m}j}=h^{ik}\cdot g_{\overline{k}%
\overline{l}}\cdot \frac{\partial p}{\partial \rho _{;l}}\cdot \rho
_{;m}\cdot h^{\overline{m}j}\,\,\,\,\,\text{.}  \label{2.6}
\end{equation}

\subsection{Invariant projections of the symmetric energy-momentum tensor of
Belinfante}

1. Rest mass density $\rho _T$. 
\begin{equation}
\rho _T:=\frac 1{e^2}\cdot \,_sT_k\,^i\cdot u_{\overline{i}}\cdot u^{%
\overline{k}}=-\frac 1{e^2}\cdot p\cdot u_{\overline{k}}\cdot u^{\overline{k}%
}\,\,\,\,\,\,\,\text{.}  \label{2.7}
\end{equation}

2. Conductive momentum density $^T\overline{\pi }$. 
\begin{equation}
^T\overline{\pi }\,^i:=\frac 1e\cdot \,T_l\,^k\cdot u_{\overline{k}}\cdot h^{%
\overline{l}i}=0\text{ \thinspace \thinspace \thinspace ,\thinspace
\thinspace \thinspace \thinspace \thinspace \thinspace \thinspace \thinspace
\thinspace \thinspace \thinspace \thinspace \thinspace \thinspace \thinspace
\thinspace \thinspace \thinspace \thinspace }T_l\,^k=0\text{\thinspace
\thinspace \thinspace \thinspace \thinspace \thinspace .}  \label{2.8}
\end{equation}

3. Conductive energy flux density $e\cdot \,^T\overline{s}$. 
\begin{equation}
^T\overline{s}^i:=\frac 1e\cdot h^{ij}\cdot g_{\overline{j}\overline{k}%
}\cdot \,T_l\,^k\cdot u^{\overline{l}}=0\,\,\,\,\,\,\,\,\,\text{,\thinspace
\thinspace \thinspace \thinspace \thinspace \thinspace \thinspace }T_l\,^k=0%
\text{\thinspace \thinspace \thinspace \thinspace \thinspace \thinspace
.\thinspace \thinspace \thinspace \thinspace \thinspace }  \label{2.9}
\end{equation}

4. Stress tensor density $^T\overline{S}$. 
\begin{equation}
^T\overline{S}^{ij}:=h^{ik}\cdot g_{\overline{k}\overline{l}}\cdot
\,T_m\,^l\cdot h^{\overline{m}j}=0\,\,\,\,\,\,\,\,\text{.}  \label{2.10}
\end{equation}

\subsection{Invariant projections of the variational energy-momentum tensor
of Euler-Lagrange}

1. Rest mass density $\rho _Q$. 
\begin{eqnarray*}
\rho _Q &=&-\frac 1{e^2}\cdot \,\overline{Q}_k\,^i\cdot u_{\overline{i}%
}\cdot u^{\overline{k}}=\frac 2{e^2}\cdot \frac{\partial p}{\partial g_{mn}}%
\cdot g_{\overline{k}m}\cdot g_n^{\underline{i}}\cdot u_{\overline{i}}\cdot
u^{\overline{k}}- \\
&&-\frac 1{e^2}\cdot \frac{\partial p}{\partial u^k}\cdot u^i\cdot u_{%
\overline{i}}\cdot u^{\overline{k}}
\end{eqnarray*}
\begin{equation}
\rho _Q=\frac 2{e^2}\cdot \frac{\partial p}{\partial g_{mn}}\cdot g_{%
\overline{k}m}\cdot u_n\cdot u^{\overline{k}}--\frac 1e\cdot \frac{\partial p%
}{\partial u^k}\cdot u^{\overline{k}}\,\,\,\text{.}  \label{2.11}
\end{equation}

2. Conductive momentum density $^Q\pi $. 
\begin{eqnarray}
^Q\pi \,^i &=&-\frac 1e\cdot \,\overline{Q}_l\,^k\cdot u_{\overline{k}}\cdot
h^{\overline{l}i}=  \notag \\
&=&-\frac 1e\cdot (\frac{\partial p}{\partial u^l}\cdot u^k\cdot u_{%
\overline{k}}\cdot h^{\overline{l}i}-2\cdot \frac{\partial p}{\partial g_{mn}%
}\cdot g_{\overline{l}m}\cdot g_n^{\underline{k}}\cdot u_{\overline{k}}\cdot
h^{\overline{l}i})=  \notag \\
&=&\frac 2e\cdot \frac{\partial p}{\partial g_{mn}}\cdot g_{\overline{l}%
m}\cdot u_n\cdot h^{\overline{l}i}-\frac{\partial p}{\partial u^l}\cdot h^{%
\overline{l}i}\text{ \thinspace \thinspace \thinspace \thinspace .}
\label{2.12}
\end{eqnarray}

3. Conductive energy flux density $e\cdot \,^Qs$. 
\begin{eqnarray*}
^Qs^i &=&-\frac 1e\cdot h^{ij}\cdot g_{\overline{j}\overline{k}}\cdot \,%
\overline{Q}_l\,^k\cdot u^{\overline{l}}= \\
&=&-\frac 1e\cdot h^{ij}\cdot g_{\overline{j}\overline{k}}\cdot u^{\overline{%
l}}\cdot (\frac{\partial p}{\partial u^l}\cdot u^k-2\cdot \frac{\partial p}{%
\partial g_{mn}}\cdot g_{\overline{l}m}\cdot g_n^{\underline{k}})= \\
&=&\frac 2e\cdot h^{ij}\cdot g_{\overline{j}\overline{k}}\cdot u^{\overline{l%
}}\cdot g_{\overline{l}m}\cdot g_n^{\underline{k}}\cdot \frac{\partial p}{%
\partial g_{mn}}-\frac 1e\cdot h^{ij}\cdot g_{\overline{j}\overline{k}}\cdot
u^{\overline{l}}\cdot u^k\cdot \frac{\partial p}{\partial u^l}\,\,\,\,\,%
\text{,}
\end{eqnarray*}
\begin{equation*}
h^{ij}\cdot g_{\overline{j}\overline{k}}\cdot u^k=0\,\,\,\,\,\,\,\text{,} 
\end{equation*}

\begin{equation}
^Qs^i=\frac 2e\cdot h^{ij}\cdot g_{\overline{j}\overline{n}}\cdot \frac{%
\partial p}{\partial g_{mn}}\cdot g_{m\overline{l}}\cdot u^{\overline{l}%
}\,\,\,\,\,\,\text{.}  \label{2.13}
\end{equation}

4. Stress tensor density $^QS$. 
\begin{eqnarray*}
^QS^{ij} &=&-h^{ik}\cdot g_{\overline{k}\overline{l}}\cdot \,\overline{Q}%
_m\,^l\cdot h^{\overline{m}j}= \\
&=&-h^{ik}\cdot g_{\overline{k}\overline{l}}\cdot (\frac{\partial p}{%
\partial u^m}\cdot u^l-2\cdot \frac{\partial p}{\partial g_{rs}}\cdot g_{%
\overline{m}r}\cdot g_s^{\underline{l}})\cdot h^{\overline{m}j}= \\
&=&-h^{ik}\cdot g_{\overline{k}\overline{l}}\cdot u^l\cdot \frac{\partial p}{%
\partial u^m}\cdot h^{\overline{m}j}+2\cdot h^{ik}\cdot g_{\overline{k}%
\overline{l}}\cdot \frac{\partial p}{\partial g_{rs}}\cdot g_{\overline{m}%
r}\cdot g_s^{\underline{l}}\cdot h^{\overline{m}j}\,\,\,\,\,\text{,}
\end{eqnarray*}
\begin{equation}
^QS^{ij}=2\cdot h^{ik}\cdot g_{\overline{k}l}\cdot \frac{\partial p}{%
\partial g_{nl}}\cdot g_{\overline{m}n}\cdot h^{\overline{m}j}\,\,\,\,\,\,%
\text{.}  \label{2.14}
\end{equation}

\subsection{Noether's identities}

From the second covariant Noether identity 
\begin{equation}
\overline{\theta }-\,_sT\equiv Q  \label{2.15}
\end{equation}
we obtain the identities:

1. $\rho _\theta \equiv \rho _T-\rho _Q:$%
\begin{equation}
\frac{\partial p}{\partial \rho _{;i}}\cdot u_{\overline{i}}\cdot u^{%
\overline{k}}\cdot \rho _{;k}\equiv e\cdot \frac{\partial p}{\partial u^k}%
\cdot u^{\overline{k}}-2\cdot \frac{\partial p}{\partial g_{mn}}\cdot g_{%
\overline{k}m}\cdot u_n\cdot u^{\overline{k}}\,\,\,\,\,\,\,\text{.}
\label{2.16}
\end{equation}

2. $^\theta \overline{\pi }\equiv \,^T\overline{\pi }-\,^Q\pi :$%
\begin{equation}
\frac{\partial p}{\partial \rho _{;k}}\cdot \rho _{;l}\cdot u_{\overline{k}%
}\cdot h^{\overline{l}i}\equiv e\cdot \frac{\partial p}{\partial u^l}\cdot
h^{\overline{l}i}-2\cdot \frac{\partial p}{\partial g_{mn}}\cdot g_{%
\overline{l}m}\cdot u_n\cdot h^{\overline{l}i}\,\,\,\,\,\,\text{.}
\label{2.17}
\end{equation}

3. $^\theta \overline{s}\equiv \,^T\overline{s}-\,^Qs:$%
\begin{equation}
h^{ij}\cdot g_{\overline{j}\overline{k}}\cdot \frac{\partial p}{\partial
\rho _{;k}}\cdot \rho _{;l}\cdot u^{\overline{l}}\equiv -2\cdot h^{ij}\cdot
g_{\overline{j}k}\cdot \frac{\partial p}{\partial g_{km}}\cdot g_{m\overline{%
l}}\cdot u^{\overline{l}}\,\,\,\,\,\,\,\text{.}  \label{2.18}
\end{equation}

4. $^\theta \overline{S}\equiv \,^T\overline{S}-\,^QS:$%
\begin{equation}
h^{ik}\cdot g_{\overline{k}\overline{l}}\cdot \frac{\partial p}{\partial
\rho _{;l}}\cdot \rho _{;m}\cdot h^{\overline{m}j}\equiv 2\cdot h^{ik}\cdot
g_{\overline{k}l}\cdot \frac{\partial p}{\partial g_{nl}}\cdot g_{\overline{m%
}n}\cdot h^{\overline{m}j}\,\,\,\,\,\,\,\text{.}  \label{2.20}
\end{equation}

\subsection{Explicit form of the energy-momentum tensors}

1. Generalized canonical energy-momentum tensor $\overline{\theta }$. 
\begin{equation}
\overline{\theta }=(\rho _\theta +\frac 1e\cdot k\cdot L)\cdot u\otimes
g(u)-p\cdot Kr+u\otimes g(^\theta \overline{\pi })+\,^\theta \overline{s}%
\,\otimes g(u)+(^\theta \overline{S})g\text{ \thinspace \thinspace
\thinspace .}  \label{2.21}
\end{equation}

2. Symmetric energy-momentum tensor of Belinfante $_sT$. 
\begin{equation}
_sT=-p\cdot Kr\,\,\,\,\,\,\text{.}  \label{2.22}
\end{equation}

3. Variational energy-momentum tensor of Euler-Lagrange $Q$. 
\begin{equation}
Q=-\rho _Q\cdot u\otimes g(u)-u\otimes g(^Q\pi )-\,^Qs\,\otimes g(u)-(^QS)g%
\text{ \thinspace \thinspace \thinspace .}  \label{2.23}
\end{equation}

The form of the symmetric energy-momentum tensor of Belinfante shows that a
fluid, described by a Lagrangian density of the type $\mathbf{L}=\sqrt{-d_g}%
\cdot p(g_{ij}$, $\rho $, $\rho _{;i}$, $u^i)$, could be considered as a
fluid in equilibrium state \cite{Loizjanskii}. Such an equilibrium state
could be related to the Euler equation of a static state of a fluid.

\section{Covariant divergencies of the energy-momentum tensors and
Navier-Stokes' equations}

\subsection{Covariant divergency}

The covariant divergency of the energy-momentum tensors could be written in
the forms respectively \cite{Manoff-2}, \cite{Manoff-4a}

1. Covariant divergency of the generalized canonical energy-momentum tensor. 
\begin{eqnarray}
\overline{g}(\delta \overline{\theta }) &=&(\rho _\theta +\frac 1e\cdot
k\cdot p)\cdot a+  \notag \\
&&+[u(\rho _\theta +\frac 1e\cdot k\cdot p)+(\rho _\theta +\frac 1e\cdot
k\cdot p)\cdot \delta u+\delta ^\theta \overline{s}]\cdot u-  \notag \\
&&-\overline{g}(Krp)-p\cdot \overline{g}(\delta Kr)+\delta u\cdot \,^\theta 
\overline{\pi }+\nabla _u^{\,\,\,\,\,\theta }\overline{\pi }+\nabla
_{^\theta \overline{s}}u+  \notag \\
&&+(\rho _\theta +\frac 1e\cdot k\cdot p)\cdot \overline{g}(\nabla _ug)(u)+%
\overline{g}(\nabla _ug)(^\theta \overline{\pi })+\overline{g}(\nabla
_{^\theta \overline{s}}g)(u)+  \notag \\
&&+\overline{g}(\delta ((^\theta \overline{S})g))\,\,\,\,\text{.}
\label{3.1}
\end{eqnarray}

2. Covariant divergency of the symmetric energy-momentum tensor of
Belinfante. 
\begin{eqnarray}
\delta \,_sT &=&-\delta (p\cdot Kr)=-(Krp+p\cdot \delta Kr)\,\,\,\,\,\,\text{%
,}  \notag \\
\overline{g}(\delta _sT) &=&-\overline{g}(Krp)-p\cdot \overline{g}(\delta
Kr)\,\,\,\,\,\,\,\,\text{.}  \label{3.2}
\end{eqnarray}

3. Covariant divergency of the variational energy-momentum tensor of
Euler-Lagrange. 
\begin{eqnarray}
\overline{g}(\delta Q) &=&-\rho _Q\cdot a-(u\rho _Q+\rho _Q\cdot \delta
u+\delta \,^Qs)\cdot u-  \notag \\
&&-\delta u\cdot \,^Q\pi -\nabla _u\,^Q\pi -\nabla _{^Qs}u-  \notag \\
&&-\rho _Q\cdot \overline{g}(\nabla _ug)(u)-\overline{g}(\nabla _ug)(^Q\pi )-%
\overline{g}(\nabla _{^Qs}g)(u)-\overline{g}(\delta ((^QS)g))\,\,\,\,\text{.}
\label{3.3}
\end{eqnarray}

\subsection{Navier-Stokes' equations}

The Navier-Stokes equations follow from the second Navier-Stokes identity 
\cite{Manoff-4a} 
\begin{equation}
h_u[\overline{g}(F)]+h_u[\overline{g}(\delta \theta )]\equiv 0\text{
\thinspace \thinspace \thinspace }  \label{3.4}
\end{equation}
in the form 
\begin{equation}
h_u[\overline{g}(\delta \theta )]=0\,\,\,\,\,\,\,\,\text{.}  \label{3.5}
\end{equation}

In its explicit form the Navier-Stokes equation could be written as 
\begin{eqnarray*}
&&(\rho _\theta +\frac 1e\cdot k\cdot p)\cdot h_u(a)-h_u[\overline{g}%
(Krp)]-p\cdot h_u[\overline{g}(\delta Kr)]+ \\
&&+\delta u\cdot h_u(\,^\theta \overline{\pi })+h_u(\nabla
_u^{\,\,\,\,\,\theta }\overline{\pi })+h_u(\nabla _{^\theta \overline{s}}u)+
\\
&&+(\rho _\theta +\frac 1e\cdot k\cdot p)\cdot h_u[\overline{g}(\nabla
_ug)(u)]+h_u[\overline{g}(\nabla _ug)(^\theta \overline{\pi })]+
\end{eqnarray*}
\begin{equation}
+h_u[\overline{g}(\nabla _{^\theta \overline{s}}g)(u)]+h_u[\overline{g}%
(\delta ((^\theta \overline{S})g))]\,=0\,\,\,\,\,\,\,\,\,\text{.}
\label{3.6}
\end{equation}

If we write the explicit form of the expressions 
\begin{eqnarray}
h_u(a) &=&g(a)-\frac 1e\cdot g(u,a)\cdot g(u)\,\,\,\,\,\text{,}  \label{3.7}
\\
h_u[\overline{g}(Krp)] &=&g[\overline{g}(Krp)]-\frac 1e\cdot g(u,\overline{g}%
(Krp))\cdot g(u)=  \notag \\
&=&Krp-\frac 1e\cdot g(u,\overline{g}(Krp))\cdot g(u)\,\,\,\,\,  \label{3.8}
\end{eqnarray}

in the Navier-Stokes equation we obtain it in the form 
\begin{eqnarray*}
&&(\rho _\theta +\frac 1e\cdot k\cdot p)\cdot g(a)-\frac 1e\cdot (\rho
_\theta +\frac 1e\cdot k\cdot p)\cdot g(u,a)\cdot g(u)- \\
&&-Krp+\frac 1e\cdot g(u,\overline{g}(Krp))\cdot g(u)-p\cdot h_u[\overline{g}%
(\delta Kr)]+ \\
&&+\delta u\cdot h_u(\,^\theta \overline{\pi })+h_u(\nabla
_u^{\,\,\,\,\,\theta }\overline{\pi })+h_u(\nabla _{^\theta \overline{s}}u)+
\\
&&+(\rho _\theta +\frac 1e\cdot k\cdot p)\cdot h_u[\overline{g}(\nabla
_ug)(u)]+h_u[\overline{g}(\nabla _ug)(^\theta \overline{\pi })]+
\end{eqnarray*}
\begin{equation}
+h_u[\overline{g}(\nabla _{^\theta \overline{s}}g)(u)]+h_u[\overline{g}%
(\delta ((^\theta \overline{S})g))]\,=0\,\,\,\,\,\,\,\,\,\text{.}
\label{3.9}
\end{equation}

After contracting the last (previous) equation with $\overline{g}$ and
introducing the abbreviation 
\begin{equation}
\overline{\rho }:=(\rho _\theta +\frac 1e\cdot k\cdot p)  \label{3.10}
\end{equation}
we find the Navier-Stokes equation in the form 
\begin{eqnarray*}
&&\overline{\rho }\cdot a-\overline{g}(Krp)-\frac 1e\cdot [\overline{\rho }%
\cdot g(u,a)-g(u,\overline{g}(Krp))]\cdot u- \\
&&-p\cdot \overline{g}(h_u)[\overline{g}(\delta Kr)]+\delta u\cdot \overline{%
g}(h_u)(\,^\theta \overline{\pi })+\overline{g}(h_u)(\nabla
_u^{\,\,\,\,\,\theta }\overline{\pi })+\overline{g}(h_u)(\nabla _{^\theta 
\overline{s}}u)+ \\
&&+\overline{\rho }\cdot \overline{g}(h_u)[\overline{g}(\nabla _ug)(u)]+%
\overline{g}(h_u)[\overline{g}(\nabla _ug)(^\theta \overline{\pi })]+
\end{eqnarray*}
\begin{equation}
+\overline{g}(h_u)[\overline{g}(\nabla _{^\theta \overline{s}}g)(u)]+%
\overline{g}(h_u)[\overline{g}(\delta ((^\theta \overline{S}%
)g))]\,=0\,\,\,\,\,\,\,\,\,\text{.}  \label{3.11}
\end{equation}

\textit{Special case:} $(\overline{L}_n,g)$-spaces. $^\theta \overline{\pi }%
:=0$, $^\theta \overline{s}:=0$, $^\theta \overline{S}:=0$. 
\begin{equation*}
\overline{\rho }\cdot a-\overline{g}(Krp)-\frac 1e\cdot [\overline{\rho }%
\cdot g(u,a)-g(u,\overline{g}(Krp))]\cdot u- 
\end{equation*}
\begin{equation}
-p\cdot \overline{g}(h_u)[\overline{g}(\delta Kr)]+\overline{\rho }\cdot 
\overline{g}(h_u)[\overline{g}(\nabla _ug)(u)]=0\,\,\,\text{.}  \label{3.12}
\end{equation}

Since $\overline{g}(h_u)\overline{g}=h^u$, we have 
\begin{equation*}
\overline{\rho }\cdot a-\overline{g}(Krp)-\frac 1e\cdot [\overline{\rho }%
\cdot g(u,a)-g(u,\overline{g}(Krp))]\cdot u- 
\end{equation*}
\begin{equation}
-p\cdot h^u(\delta Kr)+\overline{\rho }\cdot h^u[(\nabla _ug)(u)]=0\,\,\,%
\text{.}  \label{3.13}
\end{equation}

\textit{Special case:} $\overline{U}_n$- and $\overline{V}_n$-spaces. $%
\nabla _\xi g=0$ for $\forall \xi \in T(M)$, $^\theta \overline{\pi }:=0$, $%
^\theta \overline{s}:=0$, $^\theta \overline{S}:=0$. 
\begin{equation*}
\overline{\rho }\cdot a-\overline{g}(Krp)-\frac 1e\cdot [\overline{\rho }%
\cdot g(u,a)-g(u,\overline{g}(Krp))]\cdot u- 
\end{equation*}
\begin{equation}
-p\cdot h^u(\delta Kr)=0\,\,\,\text{.}  \label{3.14}
\end{equation}

\textit{Special case:} $U_n$- and $V_n$-spaces. $\nabla _\xi g=0$ for $%
\forall \xi \in T(M)$, $^\theta \overline{\pi }:=0$, $^\theta \overline{s}:=0
$, $^\theta \overline{S}:=0$.

Since $\nabla _ug=0$ and $\delta Kr=0$ we have the Navier-Stokes equation in
the form 
\begin{equation}
\overline{\rho }\cdot a-\overline{g}(Krp)-\frac 1e\cdot [\overline{\rho }%
\cdot g(u,a)-g(u,\overline{g}(Krp))]\cdot u=0\,\,\,\,\text{.}  \label{3.15}
\end{equation}

The contravariant non-null (non-isotropic) vector field $u$ is interpreted
as the velocity vector field of the moving mass elements (material points,
particles) in the fluid. The vector field $a$ is interpreted as the
acceleration of the same mass elements of the fluid. If we write the
Navier-Stokes equation in a basis (in a co-ordinate or non-co-ordinate
basis) we obtain its form 
\begin{equation}
\overline{\rho }\cdot a^i-g^{i\overline{j}}\cdot p_{,j}-\frac 1e\cdot [%
\overline{\rho }\cdot g_{\overline{k}\overline{l}}\cdot u^k\cdot a^l-u^{%
\overline{l}}\cdot p_{,l}]\cdot u^i=0\,\,\,\,\,\,\,\,\text{.}  \label{3.16}
\end{equation}

On the other side, 
\begin{eqnarray*}
g(u,a) &=&\frac 12\cdot [ue-(\nabla _ug)(u,u)]\,\,\,\,\,\,\,\text{%
,\thinspace \thinspace \thinspace \thinspace \thinspace \thinspace
\thinspace \thinspace \thinspace \thinspace \thinspace }a=\nabla _uu\text{
\thinspace \thinspace \thinspace \thinspace \thinspace ,} \\
\nabla _ue &=&\nabla _u[g(u,u)]=ue=(\nabla _ug)(u,u)+2\cdot
g(a,u)\,\,\,\,\,\,\text{.}
\end{eqnarray*}

\subsubsection{Navier-Stokes' equation in $U_n$- and $V_n$-spaces}

In a $U_n$- or $V_n$-space we obtain the Navier-Stokes equation in the form 
\begin{equation}
\overline{\rho }\cdot a^i-g^{ij}\cdot p_{,j}-\frac 1e\cdot [\frac 12\cdot 
\overline{\rho }\cdot e_{,j}\cdot u^j-u^j\cdot p_{,j}]\cdot u^i=0\text{
\thinspace \thinspace \thinspace \thinspace .}  \label{3.17}
\end{equation}

If the last term of the type of $f\cdot u$ vanishes then the expression
takes the form of the \textit{Euler equation} 
\begin{equation}
\overline{\rho }\cdot a^i=g^{ij}\cdot p_{,j}\,\,\,\,\,\,\,\text{.}
\label{3.18}
\end{equation}

The vanishing of the term $f\cdot u$ with 
\begin{equation}
f:=\frac 1e\cdot [\frac 12\cdot \overline{\rho }\cdot e_{,j}\cdot
u^j-u^j\cdot p_{,j}]  \label{3.19}
\end{equation}
could be achieved in two different ways:

(a) By additional conditions (constraints) on the vector field $u$ and on
the pressure $p$. If $e:=$ const. $\neq 0$ and $\frac{dp}{d\tau }:=0$ (for $%
u=\frac d{d\tau }$) then $f=0$. These additional conditions should be
included as constraints on the Lagrangian invariant $p$ or introduced by
definition of the pressure $p$ and the velocity $u$. The condition $e:=$
const. $\neq 0$ is usually assumed in the relativistic mechanics in $V_4$%
-spaces. $\frac{dp}{d\tau }=0$ should be in some way included in the
Lagrangian formalism for $p$. This is the reason for the statement that the
Euler equation could not be found on the basis of an unconstrained
variational principle.

(b) By changing the proper time $\tau $ as the parameter of the trajectory
(world line) $x^i(\tau )$ of a mass element. This change is not related to a
variational principle and could be performed after all steps related to the
Lagrangian formalism. The change of the proper time $\tau $ is related to
the canonical form of an auto-parallel (or geodesic in a $V_n$-space)
equation.

Let us now consider the consequences from the changing the proper time $\tau 
$ in an equation of motion of a mass element (particle) in the Newtonian and
relativistic dynamics.

\section{Invariance of dynamic characteristics under changing the proper
time of the trajectory of a mass element (particle)}

1. One interesting question is related to the invariance of physical laws
under time's transformations. The general covariance of a dynamic law under
co-ordinate transformations (diffeomorphisms of the differentiable manifolds
considered as models of space or space-time) is well known and used in
different physical theories. Let us now consider how a dynamic law changes
under a change of the time's parameter in it. A good example for the
behavior of an equation under changing the time's parameter in it is the
auto-parallel equation in spaces with affine connections and metrics. It is
well known that the change of the proper time is related to the canonical or
non-canonical form of an auto--parallel equation \cite{Manoff-7}, \cite
{Manoff-4}, \cite{Manoff-4a}. Nevertheless, the physical interpretation of a
change of the proper time is not considered in details. What is the physical
interpretation of the change of the time's parameter in Newtonian and
relativistic physics?

\subsection{Transformation of the time's parameter in the second law of the
Newtonian mechanics}

The second law in the Newton mechanics could be written in the form 
\begin{equation}
\nabla _up=F\text{ \thinspace \thinspace \thinspace , \thinspace \thinspace
\thinspace \thinspace \thinspace \thinspace \thinspace \thinspace }u=\frac
d{dt}\text{ \thinspace \thinspace \thinspace \thinspace ,\thinspace
\thinspace \thinspace \thinspace \thinspace \thinspace \thinspace \thinspace 
}p=\rho \cdot u\text{ \thinspace \thinspace \thinspace \thinspace
,\thinspace \thinspace \thinspace \thinspace \thinspace \thinspace
\thinspace \thinspace }\rho =\rho (x^k\text{, }u\text{, }t)\text{ \thinspace
\thinspace , \thinspace \thinspace }F=F(x^k\text{, }u\text{, }t)\text{%
\thinspace \thinspace \thinspace \thinspace .}  \label{4.1}
\end{equation}

If the parameter $t$ interpreted as time's parameter is changed with a new
parameter $\tau $ with 
\begin{eqnarray}
t &=&t(\tau )\text{, \thinspace \thinspace \thinspace }\tau =\tau (t)\text{%
,\thinspace \thinspace \thinspace \thinspace \thinspace \thinspace }t\text{, 
}\tau \in C^r(M)\,\,\,\text{,\thinspace \thinspace \thinspace \thinspace
\thinspace \thinspace \thinspace }r\geq 2\,\,\,\,\text{,\thinspace
\thinspace \thinspace \thinspace \thinspace \thinspace }dimM=n\,\,\,\,\,(n=3)%
\text{\thinspace \thinspace \thinspace \thinspace ,}  \label{4.2} \\
\frac{dt}{d\tau } &=&\alpha (x^k)=\alpha (x^k(\tau ))=\alpha (x^k(t))\text{
\thinspace , \thinspace }\overline{u}:=\frac d{d\tau }=\alpha \cdot u=\frac{%
dt}{d\tau }\cdot \frac d{dt}\text{ \thinspace \thinspace ,}  \label{4.3} \\
F &=&F(x^k\text{, }u\text{, }t)=F(x^k\text{ ,\thinspace \thinspace }\frac
1\alpha \cdot \overline{u}\text{ , \thinspace }t(\tau ))\,\,\,\,\text{.}
\label{4.4}
\end{eqnarray}

Then the second law, written by the use of the new time's parameter $\tau $
will have the form 
\begin{equation}
\nabla _u(\rho \cdot u)=\frac 1{\alpha ^2}\cdot \nabla _{\overline{u}}(\rho
\cdot \overline{u})-\frac 1{\alpha ^2}\cdot \rho \cdot (u\alpha )\cdot 
\overline{u}=F(x^k\text{ ,\thinspace \thinspace }\frac 1\alpha \cdot 
\overline{u}\text{ , \thinspace }t(\tau ))\text{ \thinspace \thinspace
\thinspace \thinspace .}  \label{4.5}
\end{equation}

Therefore, 
\begin{eqnarray}
\nabla _up &=&\frac 1{\alpha ^2}\cdot \nabla _{\overline{u}}\overline{p}%
-\frac 1{\alpha ^2}\cdot \rho \cdot (u\alpha )\cdot \overline{u}%
=F\,\,\,\,\,\,\text{,}  \label{4.6} \\
\nabla _{\overline{u}}\overline{p} &=&\rho \cdot (u\alpha )\cdot \overline{u}%
+\alpha ^2\cdot F=\overline{F}+\overline{f}\cdot \overline{u}\,\,\,\,\,\,%
\text{,}  \label{4.7} \\
\nabla _{\overline{u}}\overline{p} &=&\overline{f}\cdot \overline{u}+%
\overline{F}\,\,\,\,\text{.}  \label{4.8}
\end{eqnarray}
where 
\begin{equation*}
\overline{p}=\rho \cdot \overline{u}\,\,\,\,\,\text{,\thinspace \thinspace
\thinspace \thinspace \thinspace \thinspace \thinspace \thinspace \thinspace
\thinspace }\overline{f}=\rho \cdot (u\alpha )\text{ \thinspace \thinspace
\thinspace \thinspace ,\thinspace \thinspace \thinspace \thinspace
\thinspace \thinspace \thinspace }\overline{F}=\alpha ^2\cdot F\,\,\,\,\,%
\text{.} 
\end{equation*}

The change of the time's parameter $t$ in the Newton second law generates in
general an additional force proportional to the velocity $u$ (or $\overline{u%
}$). A force of this type (proportional to the velocity) is usually
interpreted as a \textit{friction force}. Only under a transformation $%
t=t(\tau )$ for which 
\begin{equation}
u\alpha =\frac 1\alpha \cdot \overline{u}\alpha =\frac 1\alpha \cdot \frac
d{d\tau }(\frac{dt}{d\tau })=\frac 1\alpha \cdot \frac{d^2t}{d\tau ^2}=0%
\text{ \thinspace \thinspace \thinspace \thinspace \thinspace ,\thinspace
\thinspace \thinspace \thinspace \thinspace \thinspace \thinspace \thinspace 
}\alpha \neq 0\,\,\,\,\,\,\text{,}  \label{4.9}
\end{equation}
the second Newton law remains invariant up to the constant function $%
(1/\alpha ^2)$%
\begin{equation}
\nabla _up=\frac 1{\alpha ^2}\cdot \nabla _{\overline{u}}\overline{p}=F\text{
\thinspace \thinspace .}  \label{4.10}
\end{equation}

The condition 
\begin{equation}
\frac{d^2t}{d\tau ^2}=0\text{ }  \label{4.11}
\end{equation}
leading to the conditions 
\begin{equation}
\frac{dt}{d\tau }=a_0=\text{ const., \thinspace \thinspace \thinspace
\thinspace \thinspace \thinspace \thinspace \thinspace \thinspace }%
t=a_0\cdot \tau +b_0\text{ \thinspace \thinspace \thinspace \thinspace
,\thinspace \thinspace \thinspace \thinspace \thinspace \thinspace
\thinspace \thinspace \thinspace \thinspace \thinspace \thinspace \thinspace 
}b_0=\text{ const.\thinspace ,}  \label{4.12}
\end{equation}
shows that only under linear transformations of the time $t=t(\tau )$ the
Newton second law is form invariant 
\begin{equation}
\nabla _up=F\,\,\,\,\,\,\text{,\thinspace \thinspace \thinspace \thinspace
\thinspace \thinspace \thinspace \thinspace \thinspace }\nabla _{\overline{u}%
}\overline{p}=\overline{F}\,\,\,\,\,\,\text{.}\,\text{\thinspace }
\label{4.13}
\end{equation}

If a mass element (particle) with rest mass density $\rho $ is a free test
particle $(F=0)$ moving under condition 
\begin{equation}
\nabla _up=0\,\,\,\,\,\text{,}  \label{4.14}
\end{equation}
then the change of the parameter $t$ generates a friction force $\alpha
\cdot \overline{f}\cdot u$ and the considered motion of the particle is no
more a free motion but a motion under a friction force proportional to its
velocity. Therefore, we can generate or remove friction forces in Newtonian
mechanics by performing an appropriate transformation of the time's
parameter $t$.

\textit{Remark}. Since $\overline{u}=\alpha \cdot u$, the Lie derivative of $%
\overline{u}$ along $u$ is $\pounds _u\overline{u}=[u,\,\overline{u}%
]=(u\alpha )\cdot u$. The dragging of $\overline{u}$ along $u$ should not
change $\overline{u}$ if we consider a free motion of a particle. This is
so, because of the equation of motion $\nabla _uu=0$ leading to $\nabla _{%
\overline{u}}u=0$ and to the condition $\nabla _{\overline{u}}\overline{u}=0$%
. Therefore, the Lie derivative $\pounds _u\overline{u}=$ $\nabla _u%
\overline{u}-\nabla _{\overline{u}}u$ in a $E_n$- or $V_n$-space ($n=3,4$)
should vanish. The Lie derivative $\pounds _u\overline{u}=-\pounds _{%
\overline{u}}u$ vanishes if and only if $u\alpha =0$ or $\overline{u}\alpha
=0$. This means that the function $\alpha $ should be a constant function,
different from zero, on trajectories with tangent vectors $u$ and $\overline{%
u}$. Under this condition $(u\alpha =0)$ the momentum of a free moving
particle is an invariant quantity $\nabla _up=\nabla _{\overline{u}}%
\overline{p}$ under changing the time's parameter of the trajectory of the
particle.

\subsection{Transformation of the time's parameter for a free moving
spinless test particle in spaces with affine connections and metrics}

1. If we consider the auto-parallel equation as the equation for description
of the motion of a free spinless mass element (test particle) with constant
rest mass density we should distinguish its two forms \cite{Manoff-7}, \cite
{Manoff-8}

(a) Canonical form 
\begin{equation}
\nabla _uu=0\text{ \thinspace \thinspace ,}  \label{4.15}
\end{equation}

(b) Non-canonical form 
\begin{equation}
\nabla _uu=f\cdot u\,\,\,\,\,\,\text{.}  \label{4.16}
\end{equation}

2. The transition from the canonical to the non-canonical form could be
realized to a transformation of the proper time (parameter) of the
trajectory of the particle.

If $u:=d/d\tau $ and $\overline{u}:=(1/\alpha )\cdot u:=d/d\overline{\tau }$
then 
\begin{eqnarray}
\nabla _uu &=&\nabla _{\alpha \cdot \overline{u}}(\alpha \cdot \overline{u}%
)=\alpha \cdot [(\overline{u}\alpha )\cdot \overline{u}+\alpha \cdot \nabla
_{\overline{u}}\overline{u}]=  \notag \\
&=&\alpha ^2\cdot \nabla _{\overline{u}}\overline{u}+(\overline{u}\alpha
)\cdot u=\alpha ^2\cdot \overline{a}+f\cdot u\,\,\,\,\,\,\text{,}
\label{4.17} \\
a &=&f\cdot u+\alpha ^2\cdot \overline{a}\,\,\,\,\,\,\text{,\thinspace
\thinspace \thinspace \thinspace \thinspace \thinspace \thinspace \thinspace 
}f=\overline{u}\alpha \,\,\,\,\,\,\text{,\thinspace \thinspace \thinspace
\thinspace \thinspace \thinspace \thinspace \thinspace \thinspace \thinspace 
}\overline{a}=\nabla _{\overline{u}}\overline{u}\,\,\,\,\text{.}
\label{4.18}
\end{eqnarray}

The acceleration $a$ after changing the parameter $\tau $ is expressed by
two terms: $f\cdot u$ (proportional to $u$) and $\alpha ^2\cdot \overline{a}%
\,$\thinspace proportional to the acceleration $\overline{a}$ defined by the
new proper time $\overline{\tau }$. The term $f\cdot u$ defines a friction
acceleration.

3. If the auto-parallel equation is given in its non-canonical form as $%
\nabla _uu=\overline{f}\cdot u$ with $u=d/d\tau $ then we can change the
parameter $\tau $ with a new parameter $\overline{\tau }:=\overline{\tau }%
(\tau )$, $\tau =\tau (\overline{\tau })$, so that 
\begin{equation}
\overline{u}=\frac d{d\overline{\tau }}=\frac 1{\overline{\alpha }}\cdot u=%
\frac{d\tau }{d\overline{\tau }}\cdot \frac d{d\tau }\text{ \thinspace
\thinspace \thinspace \thinspace \thinspace \thinspace ,\thinspace
\thinspace \thinspace \thinspace \thinspace \thinspace \thinspace \thinspace
\thinspace \thinspace \thinspace }\frac{d\tau }{d\overline{\tau }}=\frac 1{%
\overline{\alpha }}\text{ \thinspace \thinspace .}  \label{4.19}
\end{equation}

The auto-parallel equation $\nabla _uu=\overline{f}\cdot u$ could be written
in the forms 
\begin{equation}
\nabla _uu=\nabla _{\overline{\alpha }\cdot \overline{u}}(\overline{\alpha }%
\cdot \overline{u})=\overline{f}\cdot \overline{\alpha }\cdot \overline{u}%
\,\,\,\,\,\,\,\text{,}  \label{4.20}
\end{equation}
\begin{equation}
\lbrack \overline{\alpha }\cdot (\overline{u}\overline{\alpha })-\overline{f}%
\cdot \overline{\alpha }]\cdot \overline{u}+\overline{\alpha }^2\cdot \nabla
_{\overline{u}}\overline{u}=0\,\,\,\,\text{.}  \label{4.21}
\end{equation}

The function $\overline{\alpha }$ could be chosen in such a way that the
condition 
\begin{equation}
\overline{\alpha }\cdot (\overline{u}\overline{\alpha })-\overline{f}\cdot 
\overline{\alpha }=0\text{ }  \label{4.22}
\end{equation}
could be fulfilled. Since $\overline{\alpha }:\neq 0$, we have 
\begin{eqnarray}
\overline{u}\overline{\alpha }-\overline{f} &=&0:\,\overline{\alpha }=%
\overline{\alpha }_0+\int \overline{f}\cdot d\overline{\tau }\,\,\,\,\,\text{%
,\thinspace \thinspace \thinspace \thinspace \thinspace \thinspace
\thinspace \thinspace \thinspace }\overline{\alpha }_0=\text{ const.,}
\label{4.23} \\
\overline{f} &=&\overline{f}(\tau )=\overline{f}(\tau (\overline{\tau }))=%
\overline{f}(\overline{\tau })\text{ \thinspace ,}  \notag
\end{eqnarray}
or 
\begin{equation}
\overline{\alpha }=\overline{\alpha }_0\cdot exp[\int \overline{f}(\tau
)\cdot d\tau ]\,\,\,\,\,\,\,\,\text{,\thinspace \thinspace \thinspace
\thinspace \thinspace }\overline{\alpha }_0=\text{ const.\thinspace
\thinspace .\thinspace \thinspace \thinspace }  \label{4.24}
\end{equation}

If $\overline{\alpha }$ fulfills the previous condition then $\nabla _{%
\overline{u}}\overline{u}=0$ and the auto-parallel equation is represented
in its canonical form.

Therefore, an acceleration of the type of a friction acceleration
(proportional to the velocity vector field of a mass element) could be
generated or removed by the use of transformation of the proper time of the
mass element. This fact could be used for removing terms proportional to the
velocity in the structures of the energy-momentum tensors and in the
structure of Navier-Stokes' equation.

4. Let as now assume that the acceleration a of a mass element with rest
mass density $\rho $ fulfills the equation 
\begin{equation*}
\nabla _uu=a=f\cdot u\,\,\,\,\,\,\,\text{,\thinspace \thinspace \thinspace
\thinspace \thinspace \thinspace \thinspace \thinspace \thinspace \thinspace
\thinspace }f\in C^r(M)\text{ ,\thinspace \thinspace \thinspace \thinspace
\thinspace \thinspace \thinspace }r\geq 2\text{ \thinspace \thinspace
\thinspace .} 
\end{equation*}

Then the change of the momentum $p=\rho \cdot u$ of a mass element moving
with the above acceleration could be written in the form 
\begin{eqnarray}
\nabla _up &=&\nabla _u(\rho \cdot u)=(u\rho )\cdot u+\rho \cdot a=
\label{4.24a} \\
&=&(u\rho )\cdot u+\rho \cdot f\cdot u=(u\rho +f\cdot \rho )\cdot u\text{
\thinspace \thinspace .}  \notag
\end{eqnarray}

If the momentum $p$ does not change under the acceleration $a=f\cdot u$,
i.e. if 
\begin{equation}
\nabla _up=\nabla _u(\rho \cdot u)=0  \label{4.24b}
\end{equation}
then 
\begin{equation}
u\rho +f\cdot \rho =0\,\,\,\,\,\,\,\,\text{,\thinspace \thinspace \thinspace
\thinspace \thinspace \thinspace \thinspace \thinspace \thinspace \thinspace
\thinspace \thinspace \thinspace }u\neq 0\,\,\,\,\,\text{,\thinspace
\thinspace \thinspace \thinspace \thinspace \thinspace \thinspace \thinspace 
}\rho \neq 0\,\,\,\,\,\text{,}  \label{4.24c}
\end{equation}
and the rest mass density $\rho $ should obey the equation 
\begin{equation*}
\frac{u\rho }\rho =-f\text{ \thinspace \thinspace \thinspace \thinspace
,\thinspace \thinspace \thinspace \thinspace \thinspace }u(log\rho
)=-f\,\,\,\,\,\,\,\text{,\thinspace \thinspace \thinspace \thinspace
\thinspace \thinspace \thinspace \thinspace \thinspace }d(log\rho )=-f\cdot
d\tau \,\,\,\,\,\,\,\text{,} 
\end{equation*}
and therefore, 
\begin{equation}
\rho =\rho _0\cdot exp[-\int f(\tau )\cdot d\tau ]\,\,\,\,\text{, \thinspace
\thinspace \thinspace \thinspace \thinspace \thinspace \thinspace }\rho _0=%
\text{ const.}  \label{4.24d}
\end{equation}

\textit{Special case}: If $f:=\lambda :=$ const. $:\geq 0$, then 
\begin{equation}
\rho =\rho _0\cdot e^{-\lambda \cdot \tau }\text{ \thinspace \thinspace
\thinspace \thinspace \thinspace .}  \label{4.24e}
\end{equation}

\subsubsection{Law of self-decay of a system of $n$-particles in a $(%
\overline{L}_n,g)$-space}

Let us now assume as above that the rest mass density $\rho $ of a mass
element with equation of motion of the type $\nabla _uu=f\cdot u$ is a
composition of the constant rest mass density $\overline{\rho }_0$ of $n$
particles, i.e. 
\begin{equation}
\rho :=n\cdot \overline{\rho }_0\,\,\,\,\text{,\thinspace \thinspace
\thinspace \thinspace \thinspace \thinspace \thinspace \thinspace \thinspace
\thinspace \thinspace \thinspace }\overline{\rho }_0=\text{ const.\thinspace 
}  \label{4.25}
\end{equation}

Then 
\begin{eqnarray}
\nabla _u(\rho \cdot u) &=&\nabla _u(n\cdot \text{\thinspace \thinspace
\thinspace }\overline{\rho }_0\cdot u)=\text{\thinspace \thinspace
\thinspace }\overline{\rho }_0\cdot (un)\cdot u+\text{\thinspace \thinspace
\thinspace }\overline{\rho }_0\cdot n\cdot \nabla _uu=  \notag \\
&=&(un)\cdot \text{\thinspace \thinspace \thinspace }\overline{\rho }_0\cdot
u+\text{\thinspace \thinspace \thinspace }\overline{\rho }_0\cdot n\cdot
f\cdot u=  \notag \\
&=&[un+n\cdot f]\cdot \overline{\rho }_0\cdot u\text{ .}  \label{4.26}
\end{eqnarray}

If $\nabla _up=\nabla _u(\overline{\rho }_0\cdot n\cdot u)=0$, i.e. if the
momentum of the mass element with momentum $p$ and with rest mass density $%
\rho =n\cdot \overline{\rho }_0$ does not change along its trajectory (world
line), then 
\begin{equation}
un+n\cdot f=0\text{ \thinspace \thinspace \thinspace \thinspace \thinspace }
\label{4.27}
\end{equation}
and 
\begin{equation}
n=n_0\cdot exp[-\int_{\tau _0}^\tau f\cdot d\tau ]\text{ \thinspace
\thinspace \thinspace \thinspace \thinspace ,\thinspace \thinspace
\thinspace \thinspace \thinspace \thinspace \thinspace \thinspace \thinspace
\thinspace \thinspace \thinspace \thinspace }n_0=\text{ const.}  \label{4.28}
\end{equation}

The number $n$ is the number of the particles in the mass element $\rho $ at
the time $\tau $. $n_0$ is the number of the particles in the mass element $%
\rho $ at the time $\tau _0$. 
\begin{equation}
n(\tau )=n(\tau _0)\cdot e^{[-\int_{\tau _0}^\tau f\cdot d\tau ]}\,\,\,\,\,\,%
\text{.}  \label{4.29}
\end{equation}

\textit{Special case}: $f:=\lambda :=$ const. $\geq 0$. 
\begin{equation}
n(\tau )=n(\tau _0)\cdot e^{-\lambda \cdot \tau }\text{ \thinspace
\thinspace .}  \label{4.30}
\end{equation}
The last expression is exactly the law of self-decay of a system of $n$
particles in the time $\tau $. Therefore, if a mass element with rest mass
density $\rho $ (consisting of $n$ particles with rest mass density $\rho
_0=const.$) does not change its momentum density $p=\rho \cdot u$ along its
world line with tangent vector $u$ under an acceleration of the friction
type $a=f\cdot u$ with $f=\lambda =$ const. then the number $n$ of the
particles in the mass element should obey the law of self-decay 
\begin{equation}
n(\tau )=n(\tau _0)\cdot e^{-\lambda \cdot \tau }\text{ \thinspace
\thinspace ,\thinspace \thinspace \thinspace \thinspace \thinspace
\thinspace \thinspace \thinspace \thinspace \thinspace }n_0=\text{ const. ,
\thinspace \thinspace \thinspace \thinspace }\lambda =\text{ const.}
\label{4.30a}
\end{equation}

It follows that from the one side, accelerations of the type $a=f\cdot u$
could be brought to zero by a corresponding transformation of the proper
time $\tau $ and on the other side, it could be important for the
description of a self-decay law on a classical level.

The consideration of the law of self-decay on an elementary mechanical level
shows that the reparametrization of the proper time could have important for
the classical physics corollaries. The motion under friction acceleration
and preservation of the momentum of a mass element despite of this friction
acceleration (which can be removed by a proper time reparametrization) could
lead on the one side, to a self-decay of a mass element (emission of
particles), but on the other side, it could lead to absorption of particles
from the vicinity of a mass element $(f:=\overline{\lambda }:=$ const. $\leq
0)$. The number of the emitted or absorbed particles $n$ could also depend
on the proper time $\tau $ (related to the mass element $\rho $) in a more
complicated manner than on the exponential law.

\subsection{Transformation of the proper time of a mass element in the
Navier-Stokes equation in $(\overline{L}_n,g)$-spaces}

1. Now we can again consider the Navier-Stokes equation in the special case
when $^\theta \overline{\pi }:=0$, $^\theta \overline{s}:=0$, and $^\theta 
\overline{S}:=0$. It could be written in the form 
\begin{equation}
\overline{\rho }\cdot a-\overline{g}(Krp)-\overline{f}\cdot u-p\cdot
h^u(\delta Kr)+\overline{\rho }\cdot h^u[(\nabla _ug)(u)]=0\,\,\text{,}
\label{4.31}
\end{equation}
or in the form 
\begin{equation}
\overline{\rho }\cdot a-\overline{g}(Krp)-\overline{f}\cdot u-h^u[p\cdot
\delta Kr-\overline{\rho }\cdot (\nabla _ug)(u)]=0\,\,\,\,\,\text{,}
\label{4.32}
\end{equation}
where 
\begin{equation}
\overline{f}:=\frac 1e\cdot [\overline{\rho }\cdot g(u,a)-g(u,\overline{g}%
(Krp))]\,\,\,\,\text{,}  \label{4.33a}
\end{equation}

\begin{equation}
h^u[p\cdot \delta Kr-\overline{\rho }\cdot (\nabla _ug)(u)]=p\cdot
h^u(\delta Kr)-\overline{\rho }\cdot h^u[(\nabla _ug)(u)]\,\,\,\text{.}
\label{4.34}
\end{equation}

The expression 
\begin{equation}
\overline{\rho }\cdot a-\overline{f}\cdot u=\overline{\rho }\cdot (a-\frac{%
\overline{f}}{\overline{\rho }})=\overline{\rho }\cdot (a-\widetilde{f}\cdot
u)\,\,\,\,\,\,\,\,\,\,\,\text{,\thinspace \thinspace \thinspace \thinspace
\thinspace \thinspace \thinspace }\widetilde{f}=\frac{\overline{f}}{%
\overline{\rho }}\,\,\,\,\text{,\thinspace \thinspace \thinspace }
\label{4.35}
\end{equation}
could be transformed in an other expression by changing the time's parameter 
$\tau $, where $u=d/d\tau $ and $\nabla _uu=a$, with a new time's parameter $%
\overline{\tau }=\overline{\tau }(\tau )$. Then 
\begin{equation}
a-\widetilde{f}\cdot u=[\overline{\alpha }(\overline{u}\overline{\alpha })-%
\widetilde{f}\cdot \overline{\alpha }]\cdot \overline{u}+\overline{\alpha }%
^2\cdot \overline{a}\,\,\,\,\,\,\,\,\,\text{,\thinspace \thinspace
\thinspace \thinspace \thinspace \thinspace \thinspace \thinspace \thinspace
\thinspace \thinspace \thinspace }\overline{a}=\nabla _{\overline{u}}%
\overline{u}\,\,\,\,\,\,\,\,\text{.}  \label{4.36}
\end{equation}

The function $\overline{\alpha }$ (see above) could be chosen in such a way
that the condition 
\begin{equation}
\overline{u}\overline{\alpha }-\widetilde{f}=0\,\,\,\,  \label{4.37}
\end{equation}
could be fulfilled. Then 
\begin{equation}
\overline{\alpha }=\overline{\alpha }_0\cdot exp[\int \widetilde{f}(\tau
)\cdot d\tau ]\,\,\,\,\,\,\text{,\thinspace \thinspace \thinspace \thinspace
\thinspace \thinspace \thinspace \thinspace \thinspace \thinspace \thinspace
\thinspace \thinspace \thinspace \thinspace \thinspace }\overline{\alpha }_0=%
\text{ const.,}  \label{4.38}
\end{equation}
and the Navier-Stokes equation could be written in the form 
\begin{equation}
\overline{\rho }\cdot \overline{\alpha }^2\cdot \overline{a}-\overline{g}%
(Krp)-h^u[p\cdot \delta Kr-\overline{\rho }\cdot (\nabla _ug)(u)]=0\,\,\,\,\,%
\text{,\thinspace \thinspace \thinspace \thinspace \thinspace \thinspace
\thinspace \thinspace }\overline{\rho }:\neq 0\,\,\,\text{, \thinspace
\thinspace }\overline{\alpha }\neq 0\,\,\text{.}  \label{4.38a}
\end{equation}
or in the form 
\begin{equation}
\overline{a}-\frac 1{\overline{\rho }}\cdot \frac 1{\overline{\alpha }%
^2}\cdot \overline{g}(Krp)-\frac 1{\overline{\rho }}\cdot \frac 1{\overline{%
\alpha }^2}\cdot h^u[p\cdot \delta Kr-\overline{\rho }\cdot (\nabla
_ug)(u)]=0\,\,\,\,\,\,\,\text{.}  \label{4.39a}
\end{equation}

\textit{Special case}: $U_n$-and $V_n$-spaces. $S:=C$, $\nabla _\xi g:=0$
for $\forall \xi \in T(M)$. For this type of spaces the Navier-Stokes
equation has the form 
\begin{equation}
\overline{a}=\frac 1{\overline{\rho }}\cdot \frac 1{\overline{\alpha }%
^2}\cdot \overline{g}(Krp)\,\,\,\,\,\,  \label{4.40}
\end{equation}
because of $\delta Kr=0$ and $\nabla _ug=0$.

2. The metric $\overline{g}$ is an arbitrary given metric. We can now
introduce a conformal to $g$ metric tensor $\widetilde{g}$ 
\begin{equation}
\widetilde{g}:=\overline{\alpha }^2\cdot g\,\,\,\text{.}  \label{4.41}
\end{equation}

Then the corresponding contravariant metric $\widetilde{\overline{g}}$ will
have the form 
\begin{equation}
\widetilde{\overline{g}}=\overline{\alpha }^{-2}\cdot \overline{g}\text{ }
\label{4.42}
\end{equation}
because of the relations 
\begin{eqnarray*}
\widetilde{g}_{\overline{i}\overline{k}}\cdot \widetilde{g}\,^{kj} &=&%
\overline{\alpha }^2\cdot g_{\overline{i}\overline{k}}\cdot \widetilde{g}%
\,^{kj}=g_i^j\text{ \thinspace \thinspace \thinspace \thinspace ,} \\
\overline{\alpha }^2\cdot g_{\overline{i}\overline{k}}\cdot g^{il}\cdot 
\widetilde{g}\,^{kj} &=&\overline{\alpha }^2\cdot \widetilde{g}%
\,^{jl}=g^{jl}\,\,\,\,\,\text{.} \\
\widetilde{g}\,^{jl} &=&\overline{\alpha }^{-2}\cdot g^{jl}\,\,\,\,\,\,\text{%
.}
\end{eqnarray*}

With the new introduced (given) conformal to $g$ metric $\widetilde{g}$ the
Navier-Stokes equation could be written as 
\begin{equation}
\overline{a}=\frac 1{\overline{\rho }}\cdot \widetilde{\overline{g}}(Krp)%
\text{ \thinspace \thinspace ,}  \label{4.43}
\end{equation}
or in a co-ordinate basis as 
\begin{equation}
\overline{a}^i=\frac 1{\overline{\rho }}\cdot \widetilde{g}\,^{ij}\cdot
p_{,j}\,\,\,\,\,  \label{4.44}
\end{equation}
which is exactly the Euler equation in (pseudo) Riemannian spaces with or
without torsion.

3. In a 3-dimensional Euclidean space $E_3$ the Euler equation will have the
form 
\begin{equation}
\overline{a}^b=\frac 1{\overline{\rho }}\cdot \widetilde{g}\,^{bc}\cdot
p_{,c}\,\,\,\,\,\,\,\,\,\,\text{,\thinspace \thinspace \thinspace \thinspace
\thinspace \thinspace \thinspace \thinspace \thinspace \thinspace \thinspace
\thinspace \thinspace \thinspace }a,b,c=1,2,3\text{ .}  \label{4.45}
\end{equation}

If we now chose the metric $\widetilde{g}$ as the flat metric in $E_3$, i.e.
if $\widetilde{g}=\eta =\eta _{ab}\cdot dx^a.dx^b$, $\eta _{ab}=(1,1,1)$
then we obtain the classical Euler equation 
\begin{eqnarray}
\overline{a}^b &=&\frac 1{\overline{\rho }}\cdot \eta \,^{bc}\cdot
p_{,c}\,\,\,\,\,\,\,\,\text{,\thinspace \thinspace \thinspace \thinspace
\thinspace \thinspace \thinspace \thinspace \thinspace }  \label{4.46} \\
\text{\thinspace \thinspace \thinspace \thinspace \thinspace }\overline{a}^b
&=&\overline{u}^b\,_{;c}\cdot \overline{u}^c\text{ \thinspace \thinspace
\thinspace \thinspace ,\thinspace \thinspace \thinspace \thinspace
\thinspace \thinspace \thinspace \thinspace \thinspace \thinspace }\overline{%
u}^c:=\frac{dx^c}{d\overline{\tau }}=\frac{dx^c}{dt}\text{ \thinspace
\thinspace ,\thinspace \thinspace \thinspace \thinspace \thinspace
\thinspace \thinspace \thinspace }d\overline{\tau }=dt\,\,\,\text{,}
\label{4.47}
\end{eqnarray}

which could be written in a form, usually used in the classical
hydrodynamics 
\begin{equation}
\overline{u}^b\,_{,4}+\overline{u}^a\cdot \overline{u}^{b\,}\,_{,a}-\frac 1{%
\overline{\rho }}\cdot p_{,b}=0\,\,\,\,\,\,\,\,\,\text{,}  \label{4.48a}
\end{equation}
\begin{equation*}
\text{\thinspace \thinspace }\overline{u}^b\,_{,4}=\frac{\partial \overline{u%
}^b}{\partial t}\,\,\,\,\text{,\thinspace \thinspace \thinspace \thinspace
\thinspace \thinspace }\overline{u}^{b\,}\,_{,a}=\frac{\partial \overline{u}%
^b}{\partial x^a}\text{ \thinspace \thinspace \thinspace .} 
\end{equation*}

Therefore, the Navier-Stokes equation for a perfect fluid in a $3$%
-dimensional Euclidean space of the classical hydrodynamics is identical
with the Euler equation for the perfect fluid.

4. We can also write the Navier-Stokes equation in the form 
\begin{equation}
a-\widetilde{f}\cdot u-\frac 1{\overline{\rho }}\cdot \overline{g}%
(Krp)-h^u[\frac p{\overline{\rho }}\cdot \delta Kr-(\nabla _ug)(u)]=0\text{
\thinspace \thinspace \thinspace ,}  \label{1.49}
\end{equation}
where 
\begin{equation*}
\widetilde{f}=\frac 1e\cdot [g(u,a)-\frac 1{\overline{\rho }}\cdot g(u,%
\overline{g}(Krp))]\text{ \thinspace \thinspace \thinspace ,} 
\end{equation*}
\begin{equation*}
\overline{\rho }=\rho _\theta +\frac 1e\cdot k\cdot p\,\,\,\,\,\text{.} 
\end{equation*}

If we change the parameter (proper time) $\tau $ of the velocity vector
field $u=d/d\tau $ with a new parameter $\overline{\tau }=\overline{\tau }%
(\tau )$, where $\tau =\tau (\overline{\tau })$ with 
\begin{eqnarray*}
\overline{u} &=&\frac d{d\overline{\tau }}=\frac{d\tau }{d\overline{\tau }}%
\cdot \frac d{d\tau }=\frac 1{\overline{\alpha }}\cdot u\,\,\,\,\,\,\text{%
,\thinspace \thinspace \thinspace \thinspace \thinspace \thinspace } \\
\text{\thinspace \thinspace \thinspace }u &=&\overline{\alpha }\cdot 
\overline{u}\,\,\,\,\,\text{,\thinspace \thinspace \thinspace \thinspace
\thinspace \thinspace \thinspace \thinspace \thinspace }\overline{\alpha }=%
\frac{d\overline{\tau }}{d\tau }\text{ \thinspace ,\thinspace \thinspace
\thinspace \thinspace \thinspace }\overline{\alpha }\in C^r(M)\,\,\,\,\text{%
, \thinspace \thinspace \thinspace \thinspace }r\geq 1\text{ \thinspace
\thinspace ,}
\end{eqnarray*}
then 
\begin{eqnarray}
a &=&\nabla _uu=\nabla _{\overline{\alpha }\cdot \overline{u}}(\overline{%
\alpha }\cdot \overline{u})=\overline{\alpha }^2\cdot \overline{a}+(%
\overline{u}\overline{\alpha })\cdot u\,\,\,\,\text{,}  \label{4.50} \\
\overline{a} &=&\nabla _{\overline{u}}\overline{u}=\frac 1{\overline{\alpha }%
^2}\cdot [a-\frac 1{\overline{\alpha }}\cdot (u\overline{\alpha })\cdot
u]\,\,\,\text{,\thinspace \thinspace \thinspace \thinspace \thinspace
\thinspace \thinspace \thinspace \thinspace \thinspace \thinspace \thinspace
\thinspace \thinspace \thinspace }\overline{\alpha }\neq 0\,\,\,\,\,\text{.}
\label{4.51}
\end{eqnarray}

If $\overline{a}$ should be collinear to $a$ (this means no change of the
direction of the acceleration $a$ if $\tau $ is replaced with $\overline{%
\tau }$) then 
\begin{equation}
\overline{a}=\frac 1{\overline{\alpha }^2}\cdot a\text{ \thinspace
\thinspace \thinspace \thinspace \thinspace \thinspace \thinspace \thinspace
and\thinspace \thinspace \thinspace \thinspace \thinspace \thinspace
\thinspace \thinspace \thinspace \thinspace \thinspace \thinspace }u%
\overline{\alpha }=0\,\,\,\,\,\text{.}  \label{4.52}
\end{equation}

$u\overline{\alpha }=0$ means that the function $\overline{\alpha }$ should
not change along the trajectory with the parameter $\tau $.

5. Let us now again consider the transformation of the expression 
\begin{equation*}
a-\widetilde{f}\cdot u 
\end{equation*}
under the change of the proper time $\tau $. 
\begin{eqnarray}
a-\widetilde{f}\cdot u &=&\overline{\alpha }^2\cdot \overline{a}+(\overline{u%
}\overline{\alpha })\cdot u-\widetilde{f}\cdot u=  \notag \\
&=&\overline{\alpha }^2\cdot \overline{a}+[(\overline{u}\overline{\alpha })-%
\widetilde{f}]\cdot u=  \notag \\
&=&\overline{\alpha }^2\cdot \overline{a}+[\frac 1{\overline{\alpha }}\cdot
(u\overline{\alpha })-\widetilde{f}]\cdot u\,\,\,\,\text{.}  \label{4.53}
\end{eqnarray}

The last term $[\frac 1{\overline{\alpha }}\cdot (u\overline{\alpha })-%
\widetilde{f}]\cdot u\,\,$could vanish under two types of conditions:

(a) Since $\overline{\alpha }$ is until now an arbitrary given function $%
\overline{\alpha }(\tau (x^k))=\overline{\alpha }(x^k)\in C^r(M)$, $r\geq 1$%
, we could specialize its form by the use of the condition 
\begin{equation}
u\overline{\alpha }-\overline{\alpha }\cdot \widetilde{f}=0\,\,\,\,\,\,\,%
\text{,\thinspace \thinspace \thinspace \thinspace \thinspace \thinspace
\thinspace \thinspace \thinspace \thinspace \thinspace }\overline{\alpha }%
\neq 0\,\,\,\text{,}  \label{4.54}
\end{equation}
leading to the equation 
\begin{equation}
\frac{u\overline{\alpha }}{\overline{\alpha }}=\widetilde{f}\text{
\thinspace \thinspace \thinspace \thinspace \thinspace \thinspace \thinspace
,\thinspace \thinspace \thinspace \thinspace \thinspace \thinspace
\thinspace \thinspace \thinspace \thinspace }u(log\overline{\alpha })=%
\widetilde{f}\,\,\,\,\,\text{,\thinspace \thinspace \thinspace \thinspace
\thinspace \thinspace \thinspace \thinspace \thinspace \thinspace \thinspace
\thinspace }\frac d{d\tau }(log\overline{\alpha })=\widetilde{f}\,\,\,\,\,%
\text{,}  \label{4.55}
\end{equation}
with the solution 
\begin{equation}
\overline{\alpha }=\overline{\alpha }_0\cdot exp(\int \widetilde{f}\cdot
d\tau )\text{\thinspace \thinspace \thinspace \thinspace \thinspace
\thinspace \thinspace \thinspace \thinspace \thinspace ,\thinspace
\thinspace \thinspace \thinspace \thinspace \thinspace \thinspace \thinspace
\thinspace \thinspace }\overline{\alpha }_0=\text{ const.}  \label{4.56}
\end{equation}

(b) If $\overline{a}$ should be collinear to $a$ then $u\overline{\alpha }=0$
and $a-\widetilde{f}\cdot u=\overline{\alpha }^2\cdot \overline{a}$ if $%
\widetilde{f}=0$.

\textit{Remark}. If $a$ is collinear to $u$, i.e. $a:=k\cdot u$ then 
\begin{equation*}
(k-\widetilde{f})\cdot u=\overline{\alpha }^2\cdot \overline{a}+[\frac 1{%
\overline{\alpha }}\cdot (u\overline{\alpha })-\widetilde{f}]\cdot u 
\end{equation*}
and 
\begin{equation*}
\overline{\alpha }^2\cdot \overline{a}=[k-\widetilde{f}-\frac 1{\overline{%
\alpha }}\cdot (u\overline{\alpha })+\widetilde{f}]\cdot u=[k-\frac 1{%
\overline{\alpha }}\cdot (u\overline{\alpha })]\cdot u 
\end{equation*}
will be also collinear to $u$. Both equations appear as auto-parallel
equations for the vector fields $u$ and $\overline{u}$ collinear to each
other. The auto-parallel equations could be considered as a very special
type of Navier-Stokes' equation. They could be investigated separately from
the more general cases when the trajectories (world lines) of the mass
elements are not auto-parallel trajectories for the velocities vectors $u$
and $\overline{u}$.

Therefore, two possible conditions could be imposed for removing the term $%
\widetilde{f}\cdot u$ from the expression of the Navier-Stokes equation by
the use of a transformation of the proper time $\tau $

(a) $\overline{\alpha }=\overline{\alpha }_0\cdot exp(\int \widetilde{f}%
\cdot d\tau )$\thinspace ,\thinspace \thinspace \thinspace \thinspace $%
\overline{\alpha }_0=$ const.

(b) $u\overline{\alpha }=0$, \thinspace $\widetilde{f}=0$.

The condition $u\overline{\alpha }=0$ assures the collinearity of $a$ and $%
\overline{a}$ even if $\widetilde{f}\neq 0$. From physical point of view,
condition (b) is more reasonable because of the condition $u\overline{\alpha 
}=0$ leading in any case to 
\begin{equation}
\overline{a}=\frac 1{\overline{\alpha }^2}\cdot a\text{ \thinspace
\thinspace \thinspace \thinspace \thinspace \thinspace \thinspace \thinspace
\thinspace \thinspace \thinspace \thinspace \thinspace \thinspace \thinspace
or to\thinspace \thinspace \thinspace \thinspace \thinspace \thinspace
\thinspace \thinspace \thinspace \thinspace \thinspace \thinspace \thinspace
\thinspace }a=\overline{\alpha }^2\cdot \overline{a}\,\,\,\,\text{.}
\label{4.57}
\end{equation}

The condition $\widetilde{f}=0$ has its physical consequences. It could be
written in the form (for $e:\neq 0$) 
\begin{equation}
g(u,a)=\frac 1{\overline{\rho }}\cdot g(u,\overline{g}(Krp))\text{
\thinspace \thinspace \thinspace .}  \label{4.58}
\end{equation}

In a co-ordinate basis 
\begin{eqnarray}
g_{\overline{i}\overline{j}}\cdot u^i\cdot a^j &=&\frac 1{\overline{\rho }%
}\cdot g_{\overline{i}\overline{j}}\cdot u^i\cdot g^{j\overline{k}}\cdot
g_k^m\cdot p_{;m}=  \notag \\
&=&\frac 1{\overline{\rho }}\cdot u^{\overline{i}}\cdot p_{,i}\,\,\,\,\text{.%
}  \label{4.59}
\end{eqnarray}

\textit{Special case:} $U_n$- and $V_n$-spaces: $S=C$, $\nabla _\xi g=0$ for 
$\forall \xi \in T(M)$, $e:=g(u,u):=e_0=$ const.\thinspace $\neq 0$. 
\begin{eqnarray}
g(u,a) &=&\frac 12\cdot [ue-(\nabla _ug)(u,u)]=0\text{ ,}  \label{4.60} \\
u^i\cdot p_i &=&\frac{dp}{d\tau }=0\,\,\,\,\,\text{.}  \label{4.61}
\end{eqnarray}

If $\widetilde{f}=0$ then the Navier-Stokes equation (when $^\theta 
\overline{\pi }:=0$, $^\theta \overline{s}:=0$, and $^\theta \overline{S}:=0$%
) takes the form of the generalized Euler equation in a $(\overline{L}_n,g)$%
-space 
\begin{equation}
a-\frac 1{\overline{\rho }}\cdot \overline{g}(Krp)-h^u[\frac p{\overline{%
\rho }}\cdot \delta Kr-(\nabla _ug)(u)]=0\text{ \thinspace }  \label{4.62}
\end{equation}
or in a co-ordinate basis 
\begin{equation}
a^i=\frac 1{\overline{\rho }}\cdot g^{i\overline{j}}\cdot p_{,j}+h^{i%
\overline{j}}\cdot (\frac p{\overline{\rho }}\cdot
g_j^k\,_{;k}-g_{jk;l}\cdot u^l\cdot u^{\overline{k}})\,\,\,\,\,\,\,\text{.}
\label{4.63}
\end{equation}

From the previous expression for $a$ (or for $a^i$), it follows, after
contraction with the vector field $u$, the condition 
\begin{equation}
g(u,a)=\frac 1{\overline{\rho }}\cdot g(u,\overline{g}(Krp))+[g(u)](h^u)[%
\frac p{\overline{\rho }}\cdot \delta Kr-(\nabla _ug)(u)]\,\,\,\,\,\text{.}
\label{4.64}
\end{equation}

Because of $[g(u)](h^u)=0$ we obtain 
\begin{equation}
g(u,a)=\frac 1{\overline{\rho }}\cdot g(u,\overline{g}(Krp))\text{ .}
\label{4.65}
\end{equation}

This means that the condition $\widetilde{f}=0$ is automatically fulfilled
if the generalized Euler equation in a $(\overline{L}_n,g)$-space is valid.
Therefore, \textit{the condition }$\widetilde{f}=0$\textit{\ appears as a
necessary and sufficient condition for the existence of the generalized
Euler equation in a }$(\overline{L}_n,g)$\textit{-space} when $^\theta 
\overline{\pi }:=0$, $^\theta \overline{s}:=0$,\thinspace \thinspace
\thinspace $^\theta \overline{S}:=0$.

In conclusion, the Euler equation in a $(\overline{L}_n,g)$-space could be
found as a special case of the Navier-Stokes equation under the following
conditions 
\begin{eqnarray}
^\theta \overline{\pi }\,\text{\thinspace } &=&0\text{,\thinspace \thinspace
\thinspace \thinspace \thinspace \thinspace \thinspace \thinspace \thinspace
\thinspace \thinspace }^\theta \overline{s}=0\text{,\thinspace \thinspace
\thinspace \thinspace \thinspace \thinspace \thinspace \thinspace \thinspace
\thinspace \thinspace \thinspace \thinspace \thinspace \thinspace }^\theta 
\overline{S}=0\text{ \thinspace \thinspace \thinspace ,}  \label{4.66} \\
u\overline{\alpha } &=&0\text{ ,\thinspace \thinspace \thinspace \thinspace
\thinspace \thinspace }\widetilde{f}=0\,\,\,\,\,\,\,\,\text{.\thinspace
\thinspace \thinspace \thinspace \thinspace \thinspace \thinspace \thinspace
\thinspace \thinspace \thinspace \thinspace \thinspace }  \label{4.67}
\end{eqnarray}
The last two conditions are related to the form invariance of the
acceleration $a$ (within a scalar factor) with respect to the change of the
proper time $\tau $ of the world lines of the mass elements of a flow. At
the same time these conditions appear as sufficient conditions for the
existence of the Euler equation as a special case of the Navier-Stokes
equation if $^\theta \overline{\pi }:=0$,\thinspace \thinspace $^\theta 
\overline{s}:=0$,\thinspace \thinspace \thinspace $^\theta \overline{S}:=0$.

\subsection{Invariance of the projective metric $h_u$ corresponding to the
contravariant non-isotropic (non-null) vector field $u=d/d\protect\tau $
under transformation of the time parameter $\protect\tau $}

1. If the time parameter $\tau $ is transformed by the use of the
transformation $\tau =\tau (\overline{\tau })$,\thinspace $\overline{\tau }=%
\overline{\tau }(\tau )$, 
\begin{equation}
u=\frac d{d\tau }=\frac{d\overline{\tau }}{d\tau }\cdot \frac d{d\overline{%
\tau }}=\overline{\alpha }\cdot \overline{u}\text{ \thinspace \thinspace
\thinspace \thinspace \thinspace ,\thinspace \thinspace \thinspace
\thinspace \thinspace \thinspace \thinspace \thinspace \thinspace }\overline{%
\alpha }=\frac{d\overline{\tau }}{d\tau }\,\,\,\,\text{,\thinspace
\thinspace \thinspace \thinspace \thinspace \thinspace \thinspace \thinspace 
}\overline{u}=\text{\thinspace }\frac d{d\overline{\tau }}\text{ \thinspace
\thinspace \thinspace ,\thinspace }  \label{4.68}
\end{equation}
then the corresponding to $u$ projective metrics 
\begin{eqnarray*}
h_u &=&g-\frac 1e\cdot g(u)\otimes g(u)\text{ \thinspace \thinspace
\thinspace \thinspace ,\thinspace \thinspace \thinspace \thinspace
\thinspace \thinspace \thinspace \thinspace }e=g(u,u):\neq 0\,\,\,\,\,\text{,%
} \\
h^u &=&\overline{g}-\frac 1e\cdot u\otimes u\,\,\,\text{,}
\end{eqnarray*}
do not change, i.e. $h_u$ and $h^u$ are form invariant under the
transformation the time parameter $\tau $ of the vector field $u$. On this
basis we can prove the following propositions:

\textit{Proposition 1}. The corresponding to a non-isotropic (non-null)
contravariant vector field $u=d/d\tau $ projective metrics $h_u$ and $h^u$
are invariant under the transformation of the parameter $\tau $ by means of
the transformation $\tau =\tau (\overline{\tau })$ and\thinspace $\overline{%
\tau }=\overline{\tau }(\tau )$.

Proof. From the explicit form of $h_u$%
\begin{equation*}
h_u=g-\frac 1e\cdot g(u)\otimes g(u)\text{ } 
\end{equation*}
and the form of $u$ after the transformation of $\tau $%
\begin{equation*}
u=\overline{\alpha }\cdot \overline{u}\,\,\,\,\,\,\,\,\,\text{, \thinspace
\thinspace \thinspace }\tau =\tau (\overline{\tau })\,\,\,\,\,\text{%
,\thinspace \thinspace \thinspace \thinspace \thinspace \thinspace
\thinspace \thinspace \thinspace \thinspace \thinspace \thinspace \thinspace
\thinspace }\overline{\alpha }:\neq 0\,\,\text{,\thinspace \thinspace
\thinspace \thinspace }\,\,\,\text{\thinspace \thinspace \thinspace
\thinspace \thinspace \thinspace \thinspace \thinspace \thinspace \thinspace
\thinspace \thinspace \thinspace } 
\end{equation*}
it follows that 
\begin{eqnarray}
h_u &=&g-\frac 1{g(\overline{\alpha }\cdot \overline{u},\overline{\alpha }%
\cdot \overline{u})}\cdot g(\overline{\alpha }\cdot \overline{u})\otimes g(%
\overline{\alpha }\cdot \overline{u})=  \notag \\
&=&g-\frac 1{\overline{\alpha }^2}\cdot \frac 1{g(\overline{u},\overline{u}%
)}\cdot \overline{\alpha }^2\cdot g(\overline{u})\otimes g(\overline{u})= 
\notag \\
&=&g-\frac 1{g(\overline{u},\overline{u})}\cdot g(\overline{u})\otimes g(%
\overline{u})=h_{\overline{u}}\,\,\,\,\text{.}  \label{4.69}
\end{eqnarray}

Therefore, $h_u=h_{\overline{u}}$, where $e=g(u,u)=g(\overline{\alpha }\cdot 
\overline{u},\overline{\alpha }\cdot \overline{u})=\overline{\alpha }^2\cdot
g(\overline{u},\overline{u})=\overline{\alpha }^2\cdot \overline{e}$, $%
g(u)=g(\overline{\alpha }\cdot \overline{u})=\overline{\alpha }\cdot g(%
\overline{u})$.

From the explicit form of $h^u$%
\begin{equation*}
h^u=\overline{g}-\frac 1e\cdot u\otimes u\,\, 
\end{equation*}
and the form of $u$ under the transformation of $\tau =\tau (\overline{\tau }%
)$, $u=\overline{\alpha }\cdot \overline{u}\,$, \thinspace \thinspace
\thinspace $\overline{\alpha }:\neq 0$,\thinspace it follows that 
\begin{eqnarray}
h^u &=&\overline{g}-\frac 1{g(\overline{\alpha }\cdot \overline{u},\overline{%
\alpha }\cdot \overline{u})}\cdot \overline{\alpha }\cdot \overline{u}%
\otimes \overline{\alpha }\cdot \overline{u}=  \label{4.70} \\
&=&\overline{g}-\frac 1{g(\overline{u},\overline{u})}\cdot \overline{u}%
\otimes \overline{u}=h^{\overline{u}}\,\,\,\text{.}  \notag
\end{eqnarray}

Therefore, $h^u=h^{\overline{u}}$. By that $e=\overline{\alpha }^2\cdot 
\overline{e}$,\thinspace \thinspace \thinspace \thinspace $\overline{e}:=g(%
\overline{u},\overline{u})$.

\textit{Remark}. Since the vector field $\overline{u}=(1/\overline{\alpha }%
)\cdot u$ is collinear to the vector field $u$ it is obvious that the
corresponding to $u$ and $\overline{u}$ projective metrics should not change
when $u$ change to $\overline{u}$. Nevertheless this conclusion should be
proved.

\section{Perfect fluids}

\subsection{Introduction}

1. A fluid for which its thermodynamic properties and its viscosity are
considered as not important and could be neglected is called a perfect
fluid. This means that the motion of a perfect fluid does not depend on
heating and viscous processes \cite{Landau}. The model of a perfect fluid
appears as the simplest model of a fluid. A perfect fluid has no stress
tensor (the stress tensor is equal to zero) and no orthogonal to its motion
(its velocity) conductive energy flux density and conductive momentum
density \cite{Loizjanskii}.

A perfect fluid is usually defined and characterized by two different ways:

(a) By means of the Euler-Lagrange equations as equations for a fluid in
static or stationary state (Euler's equation, Bernuli's equation).

(b) By means of energy-momentum tensors with vanishing stress tensor,
conductive momentum density, and conductive energy flux density. In the
relativistic continuous media mechanics the second way is preferred. The
reason for that is that the Euler-Lagrange equations appear in general as
sufficient but not as necessary conditions for the existence of an
energy-momentum tensor of a given type.

Let us now consider the characteristics of a perfect fluid more closely.

\textit{Definition}. A perfect fluid is a continuous media with an
energy-momentum tensor $G$ of the type 
\begin{equation}
G:=(\rho _G+\frac 1e\cdot k\cdot L)\cdot u\otimes g(u)-L\cdot Kr\,\,\,\,\,%
\text{,}  \label{5.1}
\end{equation}
where 
\begin{eqnarray*}
\rho _G &=&\frac 1{e^2}\cdot [g(u)](G)(u)\,\,\text{,} \\
k &=&\frac 1e\cdot [g(u)](Kr)(u)\,\,\,\,\,\text{,} \\
Kr &=&g_j^i\cdot \partial _i\otimes dx^j\,\,\,\,\,\,\,\,\text{,\thinspace
\thinspace \thinspace \thinspace \thinspace \thinspace \thinspace \thinspace
\thinspace \thinspace }e=g(u,u):\neq 0\,\,\,\,\,\,\,\,\,\text{.}
\end{eqnarray*}

The scalar invariant $L$ is a given Lagrangian invariant, interpreted in the
case of fluid mechanics as the pressure $p$ of a fluid, i.e. 
\begin{equation}
L:=p\,\,\,\,\,\text{.}  \label{5.2}
\end{equation}

\textit{Proposition 2}. The necessary and sufficient conditions for the
existence of a perfect fluid are the conditions: 
\begin{eqnarray}
^G\overline{\pi } &=&\,^k\pi =0\,\,\,\,\,\text{,}  \notag \\
^G\overline{s} &=&\,^ks=0\,\,\,\,\,\,\,\text{,}  \label{5.3} \\
^G\overline{S} &=&\,^kS=0\,\,\,\,\,\,\,\text{.}  \notag
\end{eqnarray}

Proof: 1. Necessity. From the general representation of an energy-momentum
tensor $G$ in the form \cite{Manoff-3}, \cite{Manoff-4a} 
\begin{eqnarray*}
G &\equiv &(\rho _G+\frac 1e\cdot k\cdot L)\cdot u\otimes g(u)-L\cdot Kr\,\,+
\\
&&+u\otimes g(^k\pi )+\,^ks\otimes g(u)+(^kS)g\,\,\,\,\,\text{,}
\end{eqnarray*}
it follows that the condition 
\begin{equation}
u\otimes g(^k\pi )+\,^ks\otimes g(u)+(^kS)g\,=0\,\,\,\,\,\text{,}
\label{5.4}
\end{equation}
appears as a necessary condition for the existence of a perfect fluid (in
accordance to the definition for a perfect fluid).

After contraction of the last relation:

(a) with $g(u)$ from the left side, we obtain 
\begin{equation*}
e\cdot \,^k\pi =0\text{\thinspace \thinspace \thinspace \thinspace
,\thinspace \thinspace \thinspace \thinspace \thinspace \thinspace }^k\pi =0%
\text{ \thinspace \thinspace \thinspace \thinspace \thinspace for\thinspace
\thinspace \thinspace \thinspace }e\neq 0\,\,\,\text{,\thinspace } 
\end{equation*}
because of $g(u,^ks)=0$ and $g(u,(^kS)g)=0$.

(b) with $u$ from the right side, it follows that 
\begin{equation*}
e\cdot \,^ks=0\text{ \thinspace \thinspace ,\thinspace \thinspace \thinspace
\thinspace \thinspace \thinspace \thinspace \thinspace \thinspace \thinspace 
}^ks=0\text{ \thinspace \thinspace \thinspace \thinspace \thinspace
\thinspace for\thinspace \thinspace \thinspace \thinspace }e\neq 0\,\,\,%
\text{,\thinspace \thinspace } 
\end{equation*}
because of $g(^k\pi ,u)=0$, \thinspace $(^kS)g(u)=0$.

For \thinspace $^k\pi =0$ and $^ks=0$, it follows that $^kS=0$. Therefore,
the necessary condition is equivalent to the conditions (\ref{5.4}).

2. Sufficiency. If $^k\pi =0$, \thinspace $^ks=0$, and $^kS=0$ then 
\begin{equation*}
u\otimes g(^k\pi )+\,^ks\otimes g(u)+(^kS)g\,=0\,\, 
\end{equation*}
and the energy-momentum tensor $G$ takes the form 
\begin{equation*}
G=(\rho _G+\frac 1e\cdot k\cdot L)\cdot u\otimes g(u)-L\cdot Kr\,\,\,\,\,%
\text{.} 
\end{equation*}

The energy-momentum tensor $G$ could also be written in the form 
\begin{equation}
G=\,^kG-L\cdot Kr\text{ \thinspace .}  \label{5.5}
\end{equation}

The tensor $^kG$ is called viscous (viscosity) energy-momentum tensor and $%
^kS$ is called stress tensor.

A comparison of the last expression with the expression for $G$ of a perfect
fluid leads to the form of $^kG$ as 
\begin{equation}
^kG=(\rho _G+\frac 1e\cdot k\cdot L)\cdot u\otimes g(u)\,\,\,\,\,\text{.}
\label{5.6}
\end{equation}

After contraction from the left side with $g(u)$ and from the right side
with $u$, we obtain 
\begin{eqnarray*}
\lbrack g(u)](^kG)(u) &=&(\rho _G+\frac 1e\cdot k\cdot L)\cdot g(u,u)\cdot
g(u,u)= \\
&=&e^2\cdot (\rho _G+\frac 1e\cdot k\cdot L)=e^2\cdot \rho _{^kG}\,\,\,\,\,%
\text{,}
\end{eqnarray*}
\begin{equation*}
\lbrack g(u)](^kG)(u)=\rho _G\cdot e^2+k\cdot e\cdot L\,\,\,\,\,\text{,} 
\end{equation*}
\begin{equation}
\rho _{^kG}=\frac 1{e^2}\cdot [g(u)](^kG)(u)=\rho _G+\frac 1e\cdot k\cdot
L\,\,\,\,\text{,}  \label{5.7}
\end{equation}
\begin{equation}
L=\frac ek\cdot (\rho _{^kG}-\rho _G)\,\,\,\text{.}  \label{5.8}
\end{equation}

Therefore, every Lagrangian invariant $L$ obeying the above condition could
be used for description a perfect fluid.

If we interpret the Lagrangian invariant $L$ as the pressure $p$, i.e. if $%
L=p$, in a dynamic system the condition for $L$ appears as a state equation
for the pressure $p$ of a perfect fluid 
\begin{equation}
p=\frac ek\cdot (\rho _{^kG}-\rho _G)=\frac 1{k\cdot e}\cdot
[g(u)](^kG)(u)-\frac 1k\cdot \rho _G\cdot e\,\,\,\,\,\text{.}  \label{5.9}
\end{equation}

2. The energy-momentum tensor $G$ for a perfect fluid could also be written
in the form 
\begin{eqnarray}
G &=&(\rho _G+\frac 1e\cdot k\cdot L)\cdot u\otimes g(u)-L\cdot Kr=  \notag
\\
&=&\rho _G\cdot u\otimes g(u)+\frac 1e\cdot L\cdot \{\frac 1e\cdot
[g(u)](Kr)(u)\cdot u\otimes g(u)-Kr\}  \label{5.10}
\end{eqnarray}
because of $k=(1/e)\cdot [g(u)](Kr)(u)$.

Therefore, 
\begin{eqnarray}
G &=&\rho _G\cdot u\otimes g(u)-\frac 1e\cdot L\cdot \{Kr-\frac 1e\cdot
[g(u)](Kr)(u)\cdot u\otimes g(u)\}\,\,\,\,\,\text{,}  \notag \\
(G)\overline{g} &=&\rho _G\cdot u\otimes u-\frac 1e\cdot L\cdot \{(Kr)%
\overline{g}-\frac 1e\cdot [g(u)](Kr)(u)\cdot u\otimes u\}\,\,\,\,\text{.}
\label{5.11}
\end{eqnarray}

\textit{Special case}: $(L_n,g)$-space: $S:=C$, \thinspace $k=1$. 
\begin{equation}
(G)\overline{g}=(\rho _G+\frac 1e\cdot L)\cdot u\otimes u-L\cdot \overline{g}%
\text{ \thinspace \thinspace \thinspace \thinspace }  \label{5.12}
\end{equation}
because of 
\begin{equation}
(Kr)\overline{g}=g_j^i\cdot g^{jk}\cdot \partial _i\otimes \partial
_k=g^{ik}\cdot \partial _i\otimes \partial _k=\overline{g}\,.  \label{5.13}
\end{equation}

The tensor $(G)\overline{g}$ could be written in the form 
\begin{eqnarray}
(G)\overline{g} &=&(\rho _G+\frac 1e\cdot L)\cdot u\otimes u-L\cdot 
\overline{g}=  \notag \\
&=&\rho _G\cdot u\otimes u-L\cdot (\overline{g}-\frac 1e\cdot u\otimes u)= 
\notag \\
&=&\rho _G\cdot u\otimes u-L\cdot h^u\,\,\,\,\text{,}  \label{5.14} \\
h^u &=&\overline{g}-\frac 1e\cdot u\otimes u\,\,\,\,\,\text{.}  \notag
\end{eqnarray}

Therefore, 
\begin{equation*}
g[(G)\overline{g}]=g_{\overline{i}\overline{j}}\cdot G^{ij}=\rho _G\cdot
g[u\otimes u]-L\cdot g[h^u]\text{ \thinspace \thinspace \thinspace .} 
\end{equation*}

Since 
\begin{equation}
g[u\otimes u]=g(u,u)=e\text{ \thinspace \thinspace \thinspace \thinspace
\thinspace \thinspace ,\thinspace \thinspace \thinspace \thinspace
\thinspace \thinspace \thinspace \thinspace \thinspace \thinspace \thinspace
\thinspace \thinspace \thinspace \thinspace }g[h^u]=n-1\text{ \thinspace
\thinspace \thinspace \thinspace ,}  \label{5.15}
\end{equation}
we have for $g[(G)\overline{g}]$%
\begin{equation}
g[(G)\overline{g}]=\rho _G\cdot e-(n-1)\cdot L\,\,\,\,\,\,  \label{5.16}
\end{equation}
and 
\begin{equation}
L=\frac 1{n-1}\cdot (\rho _G\cdot e-g[(G)\overline{g}])\,\,\,\,\,\text{.}
\label{5.17}
\end{equation}

If $L=p$, i.e. if $L$ is interpreted as the pressure $p$ then $p$ could be
expressed in the form 
\begin{equation}
p=\frac 1{n-1}\cdot (\rho _G\cdot e-g[(G)\overline{g}])\,\,\,\,\,\text{.}
\label{5.18}
\end{equation}

For an energy-momentum tensor $G$ with 
\begin{equation}
g[(G)\overline{g}]=0\text{ \thinspace \thinspace \thinspace }  \label{5.19}
\end{equation}
we have 
\begin{equation}
p=\frac 1{n-1}\cdot \rho _G\cdot e\,\,\,\,\,\,\,\text{.}  \label{5.20}
\end{equation}

Therefore, if a perfect fluid is described in a $(L_n,g)$-space by means of
the corresponding for this fluid energy-momentum tensor $G$, obeying the
additional condition (having the additional property) 
\begin{equation}
g[(G)\overline{g}]=g_{ij}\cdot G^{\overline{i}\overline{j}}=g_{\overline{i}%
\overline{j}}\cdot G^{ij}=g_{ij}\cdot G^{ij}=0\,\,\,\,\,  \label{5.21}
\end{equation}
then the pressure $p$ of the perfect fluid is proportional to the rest mass
density $\rho _G$ of the corresponding energy-momentum tensor $G$.

\textit{Special case}: $(L_n,g)$-spaces, $dimM:=n:=4$, $e:=e_0:=$ const. $%
:\neq 0$, $g[(G)\overline{g}]=0$. 
\begin{equation}
p=\frac 13\cdot \rho _G\cdot e\,\,\,\,\,\,\,\text{.}  \label{5.22}
\end{equation}

If the pressure $p$ and the energy-momentum tensor $G$ are given, we could
find out the value $l_u=\,\mid u\mid \,=\,\mid g(u,u)\mid ^{1/2}$ of the
velocity vector of the perfect fluid using the form of $e=\pm l_u^2$%
\begin{equation}
e=\pm l_u^2=\frac 1{\rho _G}\cdot \{(n-1)\cdot p+g[(G)\overline{g}]\}\,\,\,\,%
\text{.}  \label{5.23}
\end{equation}

\textit{Remark}. The sign before $l_u^2$ is determined by the signature Sgn $%
g$ of the metric tensor $g$. If Sgn $g>0$ then $e=-l_u^2$. If Sgn $g<0$ then 
$e=+l_u^2$.

\textit{Special case}: $(L_n,g)$-spaces, $g[(G)\overline{g}]:=0$. 
\begin{equation}
\pm l_u^2=\frac 1{\rho _G}\cdot (n-1)\cdot p\,\,\,\,\,\,\text{,}
\label{5.24}
\end{equation}
\begin{equation}
l_u=\sqrt{(n-1)\cdot \,\mid \frac p{\rho _G}\mid }\,\,\,\,\,\,\,\,\text{.}
\label{5.25}
\end{equation}

For $p\geq 0$, $\rho _G>0$, we have 
\begin{equation}
l_u=\sqrt{(n-1)\cdot \,\frac p{\rho _G}}\,\,\,\,\,\,\,\,\text{.}
\label{5.26}
\end{equation}

\textit{Special case}: $(L_n,g)$-spaces, $g[(G)\overline{g}]:=0$, $n:=4$, $%
p\geq 0$, $\rho _G>0$. 
\begin{equation}
l_u=\sqrt{3}\cdot \,\sqrt{\frac p{\rho _G}}\,\,\,\,\,\,\text{.}  \label{5.27}
\end{equation}

For a $(\overline{L}_n,g)$-space, since $g[G(\overline{g})]=g[(G)\overline{g}%
]$, we obtain for a perfect fluid, by the use of the explicit form of $G$,
the relations: 
\begin{equation*}
g[(G)\overline{g}]=(\rho _G+\frac 1e\cdot k\cdot L)\cdot g[u\otimes
u]-p\cdot g[(Kr)\overline{g}]\text{ \thinspace \thinspace \thinspace
\thinspace .} 
\end{equation*}

Since 
\begin{eqnarray*}
g[u\otimes u] &=&g(u,u)=e\,\,\,\,\,\text{,\thinspace } \\
\text{\thinspace \thinspace \thinspace \thinspace \thinspace \thinspace
\thinspace \thinspace \thinspace \thinspace }g[(Kr)\overline{g}]\text{ }
&=&g_{\overline{i}\overline{k}}\cdot g_j^i\cdot g^{\overline{j}k}=g_{%
\overline{j}\overline{k}}\cdot g^{\overline{j}k}=g_{\overline{j}\overline{k}%
}\cdot g^{mk}\cdot f^j\,_m= \\
&=&g_j^m\cdot f^j\,_m=f^m\,_m=\overline{f}\text{ \thinspace \thinspace
\thinspace \thinspace ,}
\end{eqnarray*}
it follows for $g[(G)\overline{g}]$%
\begin{eqnarray}
g[(G)\overline{g}] &=&\rho _G\cdot e+k\cdot p-p\cdot \overline{f}=  \notag \\
&=&\rho _G\cdot e-(\overline{f}-k)\cdot p\,\,\,\,\,\,\,\,\text{.}
\label{5.28}
\end{eqnarray}

Then 
\begin{equation}
p=\frac 1{\overline{f}-k}\cdot (\rho _G\cdot e-g[(G)\overline{g}%
])\,\,\,\,\,\,\text{,}  \label{5.29}
\end{equation}
\begin{equation}
e=\frac 1{\rho _G}\cdot \{(\overline{f}-k)\cdot p+g[(G)\overline{g}%
]\}\,\,\,\,\,\,\text{.}  \label{5.30}
\end{equation}

For $(L_n,g)$-spaces $k=1$ and $\overline{f}=n$ (dim$M=n$).

\section{Lagrangian formalism for a perfect fluid in special types of $(%
\overline{L}_n,g)$-spaces and in $(L_n,g)$-spaces}

Let a Lagrangian density of the type 
\begin{equation}
\mathbf{L}=\sqrt{-d_g}\cdot p(g_{ij}\text{, }\rho \text{, }\rho _{;i}\equiv
\rho _{,i}\text{, }u^i)\text{ }  \label{6.1}
\end{equation}
with 
\begin{equation}
p:=p_0+\overline{p}_1(\rho )+a_0\cdot \rho \cdot e+\overline{p}_2(u\rho )
\label{6.2}
\end{equation}
be given, where 
\begin{eqnarray}
\overline{p}_2(u\rho ) &=&\overline{p}_2(u^k\rho _{,k})\text{ \thinspace
\thinspace \thinspace \thinspace \thinspace ,\thinspace \thinspace
\thinspace \thinspace \thinspace \thinspace \thinspace \thinspace \thinspace 
}e=g(u,u):\neq 0\,\,\,\text{,}  \label{6.3} \\
p_0 &=&\text{ const.\thinspace \thinspace \thinspace \thinspace ,\thinspace
\thinspace \thinspace \thinspace \thinspace \thinspace \thinspace \thinspace
\thinspace }a_0=\text{ const. ,}  \notag \\
\rho &=&\rho (x^k)\in \otimes ^0\,_0(M)\,\,\,\,\text{.}  \notag
\end{eqnarray}

On the basis of the method of Lagrangians with covariant derivatives (MLCD)
the whole scheme of the Lagrangian theory corresponding to the pressure $p$
as Lagrangian invariant could be found in its explicit form.

\subsection{Euler-Lagrange's equations}

\subsubsection{Euler-Lagrange's equation for the invariant function $p$}

For the invariant function $\rho $ the Euler-Lagrange equation is 
\begin{equation}
\frac{\partial p}{\partial \rho }-(\frac{\partial p}{\partial \rho _{,i}}%
)_{;i}+q_i\cdot \frac{\partial p}{\partial \rho _{,i}}=0\text{ \thinspace
\thinspace ,}  \label{6.4}
\end{equation}
where 
\begin{equation*}
q=T_{ki}\,^k-\frac 12\cdot g^{\overline{k}\overline{l}}\cdot
g_{kl;i}+g_k^l\cdot g_{l;i}^k\,\,\,\,\text{.} 
\end{equation*}

By the use of the relations 
\begin{eqnarray*}
\frac{\partial p}{\partial \rho } &=&\frac{\partial \overline{p}_1}{\partial
\rho }+a_0\cdot e\,\,\,\,\,\text{,} \\
\frac{\partial p}{\partial \rho _{,i}} &=&\frac{\partial \overline{p}%
_2(u^k\cdot \rho _{,k})}{\partial \rho _{,i}}=\frac{\partial \overline{p}_2}{%
\partial (u\rho )}\cdot \frac \partial {\partial \rho _{;i}}(u^k\cdot \rho
_k)= \\
&=&\frac{\partial \overline{p}_2}{\partial (u\rho )}\cdot u^k\cdot g_k^i=%
\frac{\partial \overline{p}_2}{\partial (u\rho )}\cdot u^i\,\,\,\,\,\text{,}
\end{eqnarray*}
\begin{equation*}
\frac{\partial \overline{p}_1}{\partial \rho }+a_0\cdot e-(\frac{\partial 
\overline{p}_2}{\partial (u\rho )}\cdot u^i)_{;i}+\frac{\partial \overline{p}%
_2}{\partial (u\rho )}\cdot q_i\cdot u^i= 
\end{equation*}
\begin{equation*}
=\frac{\partial \overline{p}_1}{\partial \rho }+a_0\cdot e-[(\frac{\partial 
\overline{p}_2}{\partial (u\rho )})_{;i}-\frac{\partial \overline{p}_2}{%
\partial (u\rho )}\cdot q_i]\cdot u^i- 
\end{equation*}
\begin{equation*}
-\frac{\partial \overline{p}_2}{\partial (u\rho )}\cdot
u^i\,_{;i}=0\,\,\,\,\,\text{,} 
\end{equation*}

the Euler-Lagrange equation for $\rho $ follows in the form 
\begin{equation}
\frac{\partial \overline{p}_1}{\partial \rho }+a_0\cdot e-[(\frac{\partial 
\overline{p}_2}{\partial (u\rho )})_{;i}-\frac{\partial \overline{p}_2}{%
\partial (u\rho )}\cdot q_i]\cdot u^i-\frac{\partial \overline{p}_2}{%
\partial (u\rho )}\cdot u^i\,_{;i}=0\,\,\,\,\,\text{.}  \label{6.5}
\end{equation}

\textit{Special case:} $\overline{p}_2:=b_0\cdot up=b_0\cdot u^i\cdot p_{,i}$%
,\thinspace \thinspace \thinspace $b_0=$ const.$\neq 0$. 
\begin{equation*}
\frac{\partial \overline{p}_2}{\partial (u\rho )}=b_0\text{ ,} 
\end{equation*}
\begin{equation}
\frac{\partial \overline{p}_1}{\partial \rho }+a_0\cdot e+b_0\cdot q_i\cdot
u^i-b_0\cdot u^i\,_{;i}=0\text{ .}  \label{6.6}
\end{equation}

\subsubsection{Euler-Lagrange's equations for the velocity vector $u$}

The Euler-Lagrange equations (ELEs) for the vector field $u$ could be found
in the form 
\begin{equation}
\frac{\partial p}{\partial u^i}=a_0\cdot \rho \cdot \frac{\partial e}{%
\partial u^i}+\frac{\partial \overline{p}_2}{\partial (u\rho )}\cdot \frac{%
\partial (u\rho )}{\partial u^i}=0\,\,\,\,\text{.}  \label{6.7}
\end{equation}

Since 
\begin{equation*}
\frac{\partial e}{\partial u^i}=2\cdot g_{\overline{i}\overline{k}}\cdot
u^k\,\,\,\,\,\text{,} 
\end{equation*}
\begin{equation*}
\frac{\partial (u\rho )}{\partial u^i}=\rho _{,i}\,\,\,\,\text{,} 
\end{equation*}
\begin{equation*}
\frac{\partial p}{\partial u^i}=2\cdot a_0\cdot \rho \cdot g_{\overline{i}%
\overline{k}}\cdot u^k+\frac{\partial \overline{p}_2}{\partial (u\rho )}%
\cdot \rho _{,i}=0\,\,\,\,\text{,} 
\end{equation*}
the ELEs for $u$ follow in the form 
\begin{equation}
2\cdot a_0\cdot \rho \cdot g_{\overline{i}\overline{k}}\cdot u^k+\frac{%
\partial \overline{p}_2}{\partial (u\rho )}\cdot \rho _{,i}=0\,\,\,\,\text{.}
\label{6.8}
\end{equation}

\textit{Special case:} $\overline{p}_2:=b_0\cdot up=b_0\cdot u^i\cdot p_{,i}$%
,\thinspace \thinspace \thinspace $b_0=$ const.$\neq 0$. 
\begin{equation*}
2\cdot a_0\cdot \rho \cdot g_{\overline{i}\overline{k}}\cdot u^k+b_0\cdot
\rho _{,i}=0\,\,\,\,\,\,\text{,} 
\end{equation*}
\begin{equation}
\rho _{,i}=-2\cdot \frac{a_0}{b_0}\cdot \rho \cdot g_{\overline{i}\overline{k%
}}\cdot u^k\,\,\,\,\,\,\text{.}  \label{6.9}
\end{equation}

\subsubsection{Euler-Lagrange's equations for the metric tensor $g$}

The ELEs for $g$ could be written in the form 
\begin{equation}
\frac{\partial p}{\partial g_{kl}}+\frac 12\cdot p\cdot g^{\overline{k}%
\overline{l}}=0\text{ \thinspace .}  \label{6.10}
\end{equation}

From the relations 
\begin{eqnarray*}
\frac{\partial p}{\partial g_{kl}} &=&a_0\cdot \rho \cdot \frac{\partial e}{%
\partial g_{kl}}=a_0\cdot \rho \cdot \frac \partial {\partial
g_{kl}}(g_{ij}\cdot u^{\overline{i}}\cdot u^{\overline{j}}) \\
&=&a_0\cdot \rho \cdot u^{\overline{i}}\cdot u^{\overline{j}}\cdot \frac{%
\partial g_{ij}}{\partial g_{kl}}\,\,\,\,\,\text{,} \\
\frac{\partial g_{ij}}{\partial g_{kl}} &=&\frac 12\frac \partial {\partial
g_{kl}}(g_{ij}+g_{ji})=\frac 12\cdot (g_i^k\cdot g_j^l+g_j^k\cdot
g_i^l)\,\,\,\text{,} \\
\frac{\partial p}{\partial g_{kl}} &=&a_0\cdot \rho \cdot u^{\overline{k}%
}\cdot u^{\overline{l}}\,\,\,\,\,\,\text{,}
\end{eqnarray*}
we obtain 
\begin{equation*}
\frac{\partial p}{\partial g_{kl}}+\frac 12\cdot p\cdot g^{\overline{k}%
\overline{l}}=a_0\cdot \rho \cdot u^{\overline{k}}\cdot u^{\overline{l}%
}+\frac 12\cdot p\cdot g^{\overline{k}\overline{l}}=0\,\,\,\,\text{,} 
\end{equation*}
\begin{equation}
g^{kl}=-2\cdot a_0\cdot \frac \rho p\cdot u^k\cdot u^l\,\,\,\,\,\,\text{%
,\thinspace \thinspace \thinspace \thinspace \thinspace \thinspace
\thinspace \thinspace \thinspace \thinspace \thinspace \thinspace \thinspace
\thinspace }p:\neq 0\,\,\,\,\,\text{.}  \label{6.11}
\end{equation}

\subsubsection{Corollaries from the Euler-Lagrange equations for the metric
tensor $g$}

From 
\begin{equation*}
a_0\cdot \rho \cdot u^{\overline{k}}\cdot u^{\overline{l}}+\frac 12\cdot
p\cdot g^{\overline{k}\overline{l}}=0 
\end{equation*}

after contraction with $g_{\overline{k}\overline{l}}$ we obtain 
\begin{equation}
a_0\cdot \rho \cdot e+\frac 12\cdot n\cdot
p=0:\,\,\,\,\,\,\,\,\,\,\,\,\,\,\,\,\,\,\,\,\,\,p=-\frac 2n\cdot a_0\cdot
\rho \cdot e\,\,\,\,\text{.}  \label{6.12}
\end{equation}

On the other side, 
\begin{equation*}
p=p_0+\overline{p}_1(\rho )+a_0\cdot \rho \cdot e+\overline{p}_2(u\rho )%
\text{ .} 
\end{equation*}

From both the expression for $p$, the condition 
\begin{equation*}
(1+\frac 2n)\cdot a_0\cdot \rho \cdot e+\overline{p}_1(\rho )+\overline{p}%
_2(u\rho )=-p_0=\text{ const., \thinspace \thinspace \thinspace \thinspace
\thinspace \thinspace \thinspace }n>0\text{ \thinspace \thinspace ,} 
\end{equation*}
follows. The metric tensor components $g^{kl}$ could be represented by means
of the components $h^{kl}$ of the projective metric $h^u$ as 
\begin{equation*}
g^{kl}=h^{kl}+\frac 1e\cdot u^k\cdot u^l\text{ } 
\end{equation*}
and on the other side as 
\begin{equation*}
g^{kl}=-2\cdot a_0\cdot \frac \rho p\cdot u^k\cdot u^l\,\,\,\,\,\text{.} 
\end{equation*}

From the last two relations, it follows the relation 
\begin{equation*}
h^{kl}=-(\frac 1e+2\cdot a_0\cdot \frac \rho p)\cdot u^k\cdot
u^l\,\,\,\,\,\, 
\end{equation*}
which contradicts to the property of $h^u$ to be orthogonal to the vector
field $u$, i.e. the explicit form of $h^{kl}$ contradicts to the properties
of $h^u$%
\begin{equation*}
h^u[g(u)]=0\,\,\,\,\,\,\text{,\thinspace \thinspace \thinspace \thinspace
\thinspace \thinspace \thinspace \thinspace \thinspace }h^{ij}\cdot g_{%
\overline{j}\overline{k}}\cdot u^k=0\text{ ,} 
\end{equation*}
because of 
\begin{eqnarray}
h^{kl}\cdot g_{\overline{l}\overline{m}}\cdot u^m &=&-(\frac 1e+2\cdot
a_0\cdot \frac \rho p)\cdot g_{\overline{l}\overline{m}}\cdot u^m\cdot
u^k\cdot u^l=  \notag \\
&=&-(\frac 1e+2\cdot a_0\cdot \frac \rho p)\cdot e\cdot u^k\neq 0\text{
\thinspace \thinspace \thinspace ,}  \label{6.13} \\
e &\neq &0\,\,\text{,\thinspace \thinspace \thinspace \thinspace \thinspace
\thinspace }\frac 1e+2\cdot a_0\cdot \frac \rho p\neq 0\,\,\,\,\text{,} 
\notag
\end{eqnarray}

\textit{Remark}. If 
\begin{equation*}
\frac 1e+2\cdot a_0\cdot \frac \rho p=0 
\end{equation*}
then 
\begin{equation*}
\frac \rho p=-\frac 1{2\cdot a_0\cdot e}\text{ \thinspace \thinspace
\thinspace \thinspace \thinspace \thinspace \thinspace \thinspace \thinspace
\thinspace \thinspace \thinspace ,\thinspace \thinspace \thinspace
\thinspace \thinspace \thinspace \thinspace \thinspace \thinspace \thinspace
\thinspace \thinspace \thinspace \thinspace \thinspace \thinspace }p=-2\cdot
a_0\cdot e\cdot \rho \,\,\,\,\,\,\text{, \thinspace \thinspace \thinspace
\thinspace \thinspace \thinspace }h^{kl}=0\text{ \thinspace .} 
\end{equation*}

The expression (\ref{6.13}) means that a projective metric $h^u$, orthogonal
to $u$, cannot exist if the ELEs for $g$ are fulfilled. Therefore, we cannot
consider the sub space, orthogonal to the vector field $u$, if the ELEs for $%
g$ are valid. From this point of view, the metric tensor field $g$ could not
be considered as dynamic field variable. It should be assumed to be given as
a non-dynamic field variable.

\subsection{Energy-momentum tensors}

\subsubsection{Generalized canonical energy-momentum tensor}

\begin{equation}
\overline{\theta }_i\,^j=\frac{\partial p}{\partial \rho _{,j}}\cdot \rho
_{,i}-p\cdot g_i^j\text{ \thinspace \thinspace .}  \label{16.14}
\end{equation}

Since 
\begin{equation*}
\frac{\partial p}{\partial \rho _{,j}}=\frac{\partial \overline{p}_2}{%
\partial (u\rho )}\cdot u^j\,\,\,\text{,} 
\end{equation*}
we have 
\begin{equation}
\overline{\theta }_i\,^j=\frac{\partial \overline{p}_2}{\partial (u\rho )}%
\cdot \rho _{,i}\cdot u^j-p\cdot g_i^j\,\,\,\,\,\text{.}  \label{16.15}
\end{equation}

\textit{Special case}: $\overline{p}_2:=b_0\cdot up=b_0\cdot u^i\cdot p_{,i}$%
,\thinspace \thinspace \thinspace $b_0=$ const.$\neq 0$. 
\begin{equation}
\overline{\theta }_i\,^j=b_0\cdot \rho _{,i}\cdot u^j-p\cdot g_i^j\,\,\,\,\,%
\text{.}  \label{16.16}
\end{equation}

If the ELEs for the vector field $u$ are fulfilled then 
\begin{equation}
\overline{\theta }_i\,^j=-2\cdot a_0\cdot \rho \cdot g_{\overline{i}%
\overline{k}}\cdot u^k\cdot u^j-p\cdot g_i^j\,\,\,\,\,\text{.}  \label{6.17}
\end{equation}

\subsubsection{Symmetric energy-momentum tensor of Belinfante}

The symmetric energy-momentum tensor of Belinfante has the simple form

\begin{equation}
_sT_i\,^j=-p\cdot g_i^j\,\,\,\,\text{.}  \label{6.18}
\end{equation}

\subsubsection{Variational energy-momentum tensor of Euler-Lagrange}

The variational energy-momentum tensor of Euler-Lagrange has the form

\begin{equation}
\overline{Q}_i\,^j=\,_v\overline{Q}_i\,^j+\,_g\overline{Q}_i\,^j\text{
\thinspace \thinspace \thinspace \thinspace \thinspace ,}  \label{6.19}
\end{equation}
\begin{eqnarray}
_v\overline{Q}_i\,^j &=&g_k^j\cdot g_i^l\cdot \frac{\delta p}{\delta u^l}%
\cdot u^k=\frac{\partial p}{\partial u^i}\cdot u^j=  \notag \\
&=&2\cdot a_0\cdot \rho \cdot g_{\overline{i}\overline{k}}\cdot u^k\cdot u^j+%
\frac{\partial \overline{p}_2}{\partial (u\rho )}\cdot \rho _{,i}\cdot
u^j\,\,\,\,\,\,\text{,}  \label{6.20}
\end{eqnarray}
\begin{eqnarray*}
_g\overline{Q}_i\,^j &=&-2\cdot \frac{\delta p}{\delta g_{\underline{j}k}}%
\cdot g_{\overline{i}k}=-2\cdot \frac{\partial p}{\partial g_{\underline{j}k}%
}\cdot g_{\overline{i}k}= \\
&=&-2\cdot a_0\cdot \rho \cdot g_{\overline{i}\overline{k}}\cdot u^k\cdot
u^j\,\,\,\,\text{.}
\end{eqnarray*}

By the use of the above relations, the variational energy-momentum tensor
could be written as 
\begin{eqnarray}
\overline{Q}_i\,^j &=&\,_v\overline{Q}_i\,^j+\,_g\overline{Q}_i\,^j=  \notag
\\
&=&2\cdot a_0\cdot \rho \cdot g_{\overline{i}\overline{k}}\cdot u^k\cdot u^j+%
\frac{\partial \overline{p}_2}{\partial (u\rho )}\cdot \rho _{,i}\cdot u^j- 
\notag \\
-2\cdot a_0\cdot \rho \cdot g_{\overline{i}\overline{k}}\cdot u^k\cdot u^j
&=&\frac{\partial \overline{p}_2}{\partial (u\rho )}\cdot \rho _{,i}\cdot
u^j\,\,\,\,\,\,\,\text{.}  \label{6.21}
\end{eqnarray}

\textit{Special case}: $\overline{p}_2:=b_0\cdot up=b_0\cdot u^i\cdot p_{,i}$%
,\thinspace \thinspace \thinspace $b_0=$ const.$\neq 0$. 
\begin{equation}
\overline{Q}_i\,^j=b_0\cdot \rho _{,i}\cdot u^j\,\,\,\,\,\text{.}
\label{6.22}
\end{equation}

If the ELEs for the vector field $u$ are fulfilled then 
\begin{equation}
\overline{Q}_i\,^j=-2\cdot a_0\cdot \rho \cdot g_{\overline{i}\overline{k}%
}\cdot u^k\cdot u^j\,\,\,\,\,\,\,\,\text{.}  \label{6.23}
\end{equation}

\subsection{Invariant projections of the energy-momentum tensors}

\subsubsection{Invariant projections of the generalized energy-momentum
tensor $\protect\theta $}

(a) Rest mass density $\rho _\theta $. 
\begin{eqnarray}
\rho _\theta &=&\frac 1{e^2}\cdot \overline{\theta }_k\,^i\cdot u_{\overline{%
i}}\cdot u^{\overline{k}}=  \notag \\
&=&\frac 1e\cdot \frac{\partial \overline{p}_2}{\partial (u\rho )}\cdot u^{%
\overline{k}}\cdot \rho _{,k}-\frac 1e\cdot k\cdot p\,\,\text{,\thinspace
\thinspace \thinspace \thinspace \thinspace \thinspace \thinspace \thinspace
\thinspace \thinspace \thinspace \thinspace \thinspace }k=\frac 1e\cdot u_{%
\overline{k}}\cdot u^{\overline{k}}\,\,\,\,\text{,}  \notag \\
\rho _\theta &=&\frac 1e\cdot [\frac{\partial \overline{p}_2}{\partial
(u\rho )}\cdot u^{\overline{k}}\cdot \rho _{,k}-k\cdot
p\,]\,\,\,\,\,\,\,\,\,\,\,\text{.}  \label{6.24}
\end{eqnarray}

\textit{Special case}: $\overline{p}_2:=b_0\cdot up=b_0\cdot u^i\cdot p_{,i}$%
,\thinspace \thinspace \thinspace $b_0=$ const.$\neq 0$. 
\begin{equation}
\rho _\theta =\frac 1e\cdot [b_0\cdot u^{\overline{k}}\cdot \rho
_{,k}-k\cdot p\,]\,\,\,\,\,\,\,\,\text{.}  \label{6.25}
\end{equation}

If the ELEs for the vector field $u$ are fulfilled then 
\begin{equation}
\rho _\theta =-\frac 1e\cdot (2\cdot a_0\cdot \rho \cdot g_{\overline{k}%
\overline{l}}\cdot u^{\overline{k}}\cdot u^l+k\cdot p)\,\,\,\,\,\text{.}
\label{6.26}
\end{equation}

\textit{Special case}: $(L_n,g)$-spaces: $S=C:k=1$. If the ELEs for $u$ are
fulfilled then 
\begin{equation}
\rho _\theta =-(2\cdot a_0\cdot \rho +\frac pe)\text{ .}  \label{6.27}
\end{equation}

(b) Conductive momentum density $^\theta \overline{\pi }$.

\begin{eqnarray}
^\theta \overline{\pi }^i &=&\frac 1e\cdot \,_k\overline{\theta }_l\,^k\cdot
u_{\overline{k}}\cdot h^{\overline{l}i}=  \notag \\
&=&\frac{\partial \overline{p}_2}{\partial (u\rho )}\cdot \rho _{,l}\cdot h^{%
\overline{l}i}\,\,\,\,\,\,\text{.}  \label{6.28}
\end{eqnarray}

\textit{Special case}: $\overline{p}_2:=b_0\cdot up=b_0\cdot u^i\cdot p_{,i}$%
,\thinspace \thinspace \thinspace $b_0=$ const.$\neq 0$. 
\begin{equation}
^\theta \overline{\pi }^i=b_0\cdot \rho _{,l}\cdot h^{\overline{l}%
i}\,\,\,\,\,\,\text{.}  \label{6.29}
\end{equation}

\textit{Special case}: $(L_n,g)$-spaces: $S=C:k=1$. 
\begin{equation}
^\theta \overline{\pi }^i=\frac{\partial \overline{p}_2}{\partial (u\rho )}%
\cdot \rho _{,l}\cdot h^{li}\,\,\,\,\,\,\,\text{.}\,  \label{6.30}
\end{equation}

If the ELEs for the vector field $u$ are fulfilled then 
\begin{eqnarray}
^\theta \overline{\pi }^i &=&-2\cdot a_0\cdot \rho \cdot g_{\overline{l}%
\overline{m}}\cdot u^m\cdot h^{\overline{l}i}=  \label{6.31} \\
&=&-2\cdot a_0\cdot \rho \cdot u_{\overline{l}}\cdot \cdot h^{\overline{l}%
i}\,\,\,\,\,\,\text{.}  \notag
\end{eqnarray}

\textit{Special case}: $(L_n,g)$-spaces: $S=C:k=1$. If the ELEs for $u$ are
fulfilled then 
\begin{equation}
^\theta \overline{\pi }^i=-2\cdot a_0\cdot \rho \cdot g_{lm}\cdot u^m\cdot
h^{li}=0\,\,\,\,\,\,\text{.}  \label{6.32}
\end{equation}

Therefore, in a $(L_n,g)$-space the conductive momentum density $^\theta \pi 
$ vanishes if the ELEs for the vector field $u$ are fulfilled.

(c) Conductive energy flux density $^\theta \overline{s}$. 
\begin{eqnarray}
^\theta \overline{s}^i &=&\frac 1e\cdot h^{ij}\cdot g_{\overline{j}\overline{%
k}}\cdot \,_k\overline{\theta }_l\,^k\cdot u^{\overline{l}}=  \label{6.33} \\
&=&\frac 1e\cdot \frac{\partial \overline{p}_2}{\partial (u\rho )}\cdot \rho
_{,l}\cdot u^{\overline{l}}\cdot h^{ij}\cdot g_{\overline{j}\overline{k}%
}\cdot u^k=0\,\,\,\,\text{.}  \notag
\end{eqnarray}

The conductive energy flux density $^\theta \overline{s}$ is equal to zero
for the given Lagrangain density.

(d) Stress tensor $^\theta \overline{S}$. 
\begin{eqnarray}
^\theta \overline{S}\,^{ij} &=&h^{ik}\cdot g_{\overline{k}\overline{l}}\cdot
\,_k\overline{\theta }_m\,^l\cdot h^{\overline{m}j}=  \notag \\
&=&\frac{\partial \overline{p}_2}{\partial (u\rho )}\cdot \rho _{,m}\cdot
u^l\cdot h^{ik}\cdot g_{\overline{k}\overline{l}}\cdot h^{\overline{m}j}= 
\notag \\
&=&\frac{\partial \overline{p}_2}{\partial (u\rho )}\cdot \rho _{,m}\cdot h^{%
\overline{m}j}\cdot h^{ik}\cdot g_{\overline{k}\overline{l}}\cdot
u^l=0\,\,\,\,\,\,\text{.}  \label{6.34}
\end{eqnarray}

The stress tensor $^\theta \overline{S}$ is equal to zero for the given
Lagrangian density.

\subsubsection{Invariant projections of the symmetric energy-momentum tensor
of Belinfante $_sT$}

(a) Rest mass density $\rho _T$. 
\begin{eqnarray}
\rho _T &=&\frac 1{e^2}\cdot g_{\overline{i}\overline{j}}\cdot u^j\cdot
\,_sT_k\,^i\cdot u^{\overline{k}}=-\frac 1{e^2}\cdot p\cdot g_{\overline{i}%
\overline{j}}\cdot u^j\cdot \cdot u^{\overline{k}}\cdot g_k^i=  \notag \\
&=&-\frac 1e\cdot k\cdot p\,\,\,\,\,\text{.}  \label{6.35}
\end{eqnarray}

\textit{Special case}: $(L_n,g)$-spaces: $S=C:k=1$. 
\begin{equation}
\rho _T=-\frac 1e\cdot p\,\,\,\,\text{.}  \label{6.36}
\end{equation}

(b) Conductive momentum density $^T\overline{\pi }$. 
\begin{equation}
^T\overline{\pi }^i=\frac 1e\cdot \,_{sk}T_l\,^k\cdot u_{\overline{k}}\cdot
h^{\overline{l}i}=0\,\text{\thinspace \thinspace \thinspace \thinspace
\thinspace \thinspace ,\thinspace \thinspace \thinspace \thinspace
\thinspace \thinspace \thinspace \thinspace \thinspace \thinspace \thinspace 
}_{sk}T_l\,^k=0\text{ \thinspace \thinspace .}  \label{6.37}
\end{equation}

(c) Conductive energy flux density $^T\overline{s}$. 
\begin{equation}
^T\overline{s}^i=\frac 1e\cdot h^{ij}\cdot g_{\overline{j}\overline{k}}\cdot
\,_{sk}T_l\,^k\cdot u^{\overline{l}}=0\,\,\,\,\,\,\,\,\text{,\thinspace
\thinspace \thinspace \thinspace \thinspace \thinspace \thinspace \thinspace
\thinspace \thinspace }_{sk}T_l\,^k=0\text{ \thinspace \thinspace
.\thinspace \thinspace }  \label{6.38}
\end{equation}

(d) Stress tensor $^T\overline{S}$. 
\begin{equation}
^T\overline{S}\,^{ij}=h^{ik}\cdot g_{\overline{k}\overline{l}}\cdot
\,_{sk}T_m\,^l\cdot h^{\overline{m}j}=0\,\,\,\,\,\,\,\,\,\,\text{,\thinspace
\thinspace \thinspace \thinspace \thinspace \thinspace \thinspace \thinspace
\thinspace \thinspace \thinspace \thinspace \thinspace \thinspace \thinspace 
}_{sk}T_l\,^k=0\text{ \thinspace \thinspace .\thinspace \thinspace
\thinspace \thinspace \thinspace }  \label{6.39}
\end{equation}

\subsubsection{Invariant projections of the variational energy-momentum
tensor of Euler-Lagrange}

(a) Rest mass density $\rho _Q$. 
\begin{eqnarray}
\rho _Q &=&-\frac 1{e^2}\cdot g_{\overline{i}\overline{j}}\cdot u^j\cdot \,%
\overline{Q}_k\,^i\cdot u^{\overline{k}}=-\frac 1{e^2}\cdot \frac{\partial 
\overline{p}_2}{\partial (u\rho )}\cdot \rho _{,k}\cdot u^i\cdot u_{%
\overline{i}}\cdot u^{\overline{k}}=  \notag \\
&=&-\frac 1e\cdot \frac{\partial \overline{p}_2}{\partial (u\rho )}\cdot
\rho _{,k}\cdot u^{\overline{k}}\,\,\,\,\,\text{.}  \label{6.40}
\end{eqnarray}

\textit{Special case}: $\overline{p}_2:=b_0\cdot up=b_0\cdot u^i\cdot p_{,i}$%
,\thinspace \thinspace \thinspace $b_0=$ const.$\neq 0$. 
\begin{equation}
\rho _Q=-\frac 1e\cdot b_0\cdot \rho _{,k}\cdot u^{\overline{k}}\,\,\,\,\,%
\text{.}  \label{6.41}
\end{equation}

If the ELEs for the vector field $u$ are fulfilled then 
\begin{eqnarray}
\rho _Q &=&-\frac 1e\cdot (-2\cdot a_0\cdot \rho \cdot g_{\overline{k}%
\overline{l}}\cdot u^l)\cdot u^{\overline{k}}=  \notag \\
&=&\frac 2e\cdot a_0\cdot \rho \cdot g_{\overline{k}\overline{l}}\cdot
u^l\cdot u^{\overline{k}}\,\,\,\text{.}  \label{6.42}
\end{eqnarray}

\textit{Special case}: $(L_n,g)$-spaces: $S=C:k=1$. If the ELEs for $u$ are
fulfilled then 
\begin{equation}
\rho _Q=2\cdot a_0\cdot \rho \,\,\,\,\,\text{.}  \label{6.43}
\end{equation}

For $a_0:=\frac 12$ the rest mass density $\rho _Q$ is identical to the
introduced invariant function $\rho $ in the Lagrangian invariant $L=p$.

(b) Conductive momentum density $^Q\pi $. 
\begin{eqnarray}
^Q\pi ^i &=&-\frac 1e\cdot \,\overline{Q}_l\,^k\cdot u_{\overline{k}}\cdot
h^{\overline{l}i}=-\frac 1e\cdot \frac{\partial \overline{p}_2}{\partial
(u\rho )}\cdot \rho _{,l}\cdot u^k\cdot u_{\overline{k}}\cdot h^{\overline{l}%
i}=  \notag \\
&=&-\frac{\partial \overline{p}_2}{\partial (u\rho )}\cdot \rho _{,l}\cdot
h^{\overline{l}i}\,\,\,\,\,\text{.}  \label{6.44}
\end{eqnarray}

\textit{Special case}: $\overline{p}_2:=b_0\cdot up=b_0\cdot u^i\cdot p_{,i}$%
,\thinspace \thinspace \thinspace $b_0=$ const.$\neq 0$. 
\begin{equation}
^Q\pi ^i=-b_0\cdot \rho _{,l}\cdot h^{\overline{l}i}\,\,\,\,\,\text{.}
\label{6.45}
\end{equation}

If the ELEs for the vector field $u$ are fulfilled then 
\begin{equation}
^Q\pi ^i=2\cdot a_0\cdot \rho \cdot g_{\overline{l}\overline{m}}\cdot
u^m\cdot h^{\overline{l}i}\,\,\,\,\,\,\text{.}  \label{6.46}
\end{equation}

\textit{Special case}: $(L_n,g)$-spaces: $S=C:k=1$. If the ELEs for $u$ are
fulfilled then 
\begin{equation}
^Q\pi ^i=2\cdot a_0\cdot \rho \cdot g_{lm}\cdot u^m\cdot h^{li}=0\,\,\,\,\,%
\text{.}  \label{6.47}
\end{equation}

(c) Conductive energy flux density $^Qs$. 
\begin{eqnarray}
^Qs^i &=&-\frac 1e\cdot h^{ij}\cdot g_{\overline{j}\overline{k}}\cdot \,%
\overline{Q}_l\,^k\cdot u^{\overline{l}}=  \notag \\
&=&-\frac 1e\cdot \frac{\partial \overline{p}_2}{\partial (u\rho )}\cdot
\rho _{,l}\cdot u^k\cdot h^{ij}\cdot g_{\overline{j}\overline{k}}\cdot u^{%
\overline{l}}=0\,\,\,\,\,\,\text{.}  \label{6.48}
\end{eqnarray}

(d) Stress tensor $^QS$. 
\begin{eqnarray}
^QS^{ij} &=&-h^{ik}\cdot g_{\overline{k}\overline{l}}\cdot \,\overline{Q}%
_m\,^l\cdot h^{\overline{m}j}=  \notag \\
&=&-\,h^{ik}\cdot g_{\overline{k}\overline{l}}\cdot \frac{\partial \overline{%
p}_2}{\partial (u\rho )}\cdot \rho _{,m}\cdot u^l\cdot h^{\overline{m}j}= 
\notag \\
&=&-\frac{\partial \overline{p}_2}{\partial (u\rho )}\cdot \rho _{,m}\cdot
h^{\overline{m}j}\cdot h^{ik}\cdot g_{\overline{k}\overline{l}}\cdot
u^l=0\,\,\,\,\,\,\,\,\text{.}  \label{6.49}
\end{eqnarray}

We can summarize the results for the projective quantities if the ELEs for
the vector field $u$ are fulfilled in a $(\overline{L}_n,g)$-space and in a $%
(L_n,g)$-space in the following table

\begin{center}
\begin{tabular}{cccccccccccccc}
Quantity & $\rho _\theta $ & $^\theta \overline{\pi }$ & $^\theta \overline{s%
}$ & $^\theta \overline{S}$ & $\rho _T$ & $^T\overline{\pi }$ & $^T\overline{%
s}$ & $^T\overline{S}$ & $\rho _Q$ & $^Q\pi $ & $^Qs$ & $^QS$ &  \\ 
Space &  &  &  &  &  &  &  &  &  &  &  &  &  \\ 
$(\overline{L}_n,g)$-space & $\neq 0$ & $\neq 0$ & $0$ & $0$ & $\neq 0$ & $0$
& $0$ & $0$ & $\neq 0$ & $\neq 0$ & $0$ & $0$ &  \\ 
$(L_n,g)$-space & $\neq 0$ & $0$ & $0$ & $0$ & $\neq 0$ & $0$ & $0$ & $0$ & $%
\neq 0$ & $0$ & $0$ & $0$ &  \\ 
&  &  &  &  &  &  &  &  &  &  &  &  & 
\end{tabular}
\end{center}

It is obvious from the table that a perfect fluid described by the use of
the given Lagrangian density $\mathbf{L}$ could exist only in a $(L_n,g)$%
-space and, in general, it could not exist in a $(\overline{L}_n,g)$-space.
In special cases of $(\overline{L}_n,g)$-spaces with conformal contraction
operator $S$ with $S(dx^i,\partial _j)=f^i\,_j:=\varphi (x^k)\cdot g_j^i$
there are analogous relations as in $(L_n,g)$-spaces. In a $(L_n,g)$-space: 
\begin{eqnarray}
\theta &=&(\rho _\theta +\frac 1e\cdot p)\cdot u\otimes g(u)-p\cdot Kr\text{
, \thinspace \thinspace \thinspace \thinspace \thinspace \thinspace
\thinspace }\rho _\theta =-(2\cdot a_0\cdot \rho +\frac pe)\,\,\,\,\,\text{,}
\label{6.50} \\
_sT &=&-p\cdot Kr\,\,\,\,\,\,\text{,\thinspace \thinspace \thinspace
\thinspace \thinspace \thinspace \thinspace \thinspace \thinspace \thinspace
\thinspace \thinspace }\rho _T=-\frac pe\,\,\,\,\,\,\text{,}  \label{6.51} \\
Q &=&-\rho _Q\cdot u\otimes g(u)\,\,\,\,\,\,\,\,\text{,\thinspace \thinspace
\thinspace \thinspace \thinspace \thinspace \thinspace \thinspace \thinspace
\thinspace \thinspace \thinspace \thinspace \thinspace \thinspace \thinspace
\thinspace }\rho _Q=2\cdot a_0\cdot \rho \text{\thinspace \thinspace
\thinspace \thinspace \thinspace \thinspace ,}  \label{6.52}
\end{eqnarray}
\begin{eqnarray}
\theta &=&-2\cdot a_0\cdot \rho \cdot u\otimes g(u)-p\cdot Kr\,\,\,\,\,\text{%
,}  \label{6.53} \\
_sT &=&-p\cdot Kr\,\,\,\,\text{,}  \label{6.54} \\
Q &=&-2\cdot a_0\cdot \rho \cdot u\otimes g(u)\,\,\text{.}  \label{6.55}
\end{eqnarray}

If we chose the arbitrary constant $a_0\neq 0$ as $a_0:=-\frac 12$ then 
\begin{eqnarray}
\rho _\theta &=&\rho -\frac pe\,\,\,\,\text{,}  \label{6.56} \\
\rho _T &=&-\frac pe\,\,\,\,\,\,\,\text{,}  \label{6.57} \\
\rho _Q &=&-\rho \,\,\,\,\,\,\text{.}  \label{6.58}
\end{eqnarray}

The scalar invariant function $\rho $ is proportional to the rest mass
density $\rho _Q$ of the variational energy-momentum tensor of
Euler-Lagrange. The rest mass density $\rho _T$ of the symmetric
energy-momentum tensor of Belinfante is equal up to a sign to the rest mass
density generated by the pressure $p$ of the system. The rest mass density $%
\rho _\theta $ corresponding to the generalized canonical energy-momentum
tensor is equal to the difference of the rest mass $\rho $ and $\frac pe$\
(if $a_0=-\frac 12$).

The generalized canonical energy-momentum tensor could be considered as the
sum of the symmetric energy-momentum tensor of Belinfante and the
variational energy-momentum tensor of Euler-Lagrange 
\begin{equation}
\theta \equiv \,_sT+Q\,\,\,\,\text{.}  \label{6.59}
\end{equation}

The last relation represents the second covariant Noether identity for a
perfect fluid.

In the further considerations we will assume the existence of a perfect
fluid in a $(L_n,g)$-space or in its special cases.

\section{Navier-Stokes' and Euler's equations in $(L_n,g)$-spaces}

\subsection{Navier-Stokes' equations}

In a $(L_n,g)$-space $(S=C)$, if the Euler-Lagrange equations for the scalar
function $\rho $ and for the vector field $u$ are valid, we have the
relations for a perfect fluid in a co-ordinate basis: 
\begin{eqnarray}
g^{ik}\cdot \overline{\theta }_k\,^j\,_{;j} &=&(\rho _\theta +\frac 1e\cdot
p)\cdot a^i+  \notag \\
&&+[(\rho _\theta +\frac 1e\cdot p)_{,j}\cdot u^j+(\rho _\theta +\frac
1e\cdot p)\cdot u^j\,_{;j}]\cdot u^i-  \notag \\
&&-p_{,j}\cdot g^{ij}+(\rho _\theta +\frac 1e\cdot p)\cdot g^{il}\cdot
g_{lj;k}\cdot u^j\cdot u^k\,\,\,\,\,\,\,\text{.}  \label{7.1}
\end{eqnarray}

Since 
\begin{equation*}
\rho _\theta =-(2\cdot a_o\cdot \rho +\frac 1e\cdot p)\,\,\,\,\,\,\text{,} 
\end{equation*}
we have further 
\begin{eqnarray}
g^{ik}\cdot \overline{\theta }_k\,^j\,_{;j} &=&-2\cdot a_o\cdot \rho \cdot
a^i-  \notag \\
&&-2\cdot a_0\cdot (\rho _{,j}\cdot u^j+\rho \cdot u^j\,_{;j})\cdot
u^i-p_{,j}\cdot g^{ij}-  \notag \\
&&-2\cdot a_0\cdot \rho \cdot g^{il}\cdot g_{lj;k}\cdot u^j\cdot
u^k\,\,\,\,\,\text{.}  \label{7.2}
\end{eqnarray}

The Navier-Stokes equation could be written in the forms 
\begin{eqnarray}
\text{(a)\thinspace \thinspace \thinspace \thinspace \thinspace \thinspace
\thinspace \thinspace \thinspace \thinspace \thinspace \thinspace }h_u[%
\overline{g}(\delta \theta )] &=&0\,\,\,\,\,\,\,\text{,}  \label{7.3} \\
\text{(b)\thinspace \thinspace \thinspace \thinspace \thinspace \thinspace
\thinspace \thinspace \thinspace \thinspace \thinspace \thinspace \thinspace
\thinspace }h_{li}\cdot g^{ik}\cdot \overline{\theta }_k\,^j\,_{;j} &=&0
\label{7.4}
\end{eqnarray}

Since (see the above consideration of Navier-Stokes' equation) 
\begin{equation}
\overline{\rho }=\rho _\theta +\frac 1e\cdot p=-2\cdot a_0\cdot \rho \text{
\thinspace \thinspace \thinspace \thinspace \thinspace \thinspace
,\thinspace \thinspace \thinspace \thinspace \thinspace \thinspace
\thinspace \thinspace \thinspace \thinspace \thinspace \thinspace \thinspace
\thinspace }k=1\,\,\,\,\,\,\,\text{,\thinspace \thinspace \thinspace
\thinspace \thinspace \thinspace \thinspace \thinspace \thinspace \thinspace
\thinspace \thinspace \thinspace }\delta Kr=0\,\,\,\,\,\,\,\,\,\text{,}%
\,\,\,\,\,\,  \label{7.5}
\end{equation}
the Navier-Stokes equation for a perfect fluid in a $(L_n,g)$-space could
also be written in the forms [see the special case of $(\overline{L}_n,g)$%
-spaces with $^G\overline{\pi }:=\,^k\pi =0\,$, \thinspace $^G\overline{s}%
:=\,^ks=0\,$,\thinspace $^G\overline{S}:=\,^kS=0$.] 
\begin{equation*}
-2\cdot a_0\cdot \rho \text{ }\cdot a-\overline{g}(Krp)+\frac 1e\cdot
[2\cdot a_0\cdot \rho \cdot g(u,a)+g(u,\overline{g}(Krp))]\cdot u- 
\end{equation*}
\begin{equation}
-2\cdot a_0\cdot \rho \text{ }\cdot h^u[(\nabla _ug)(u)]=0\,\,\,\,\,\,\,\,\,%
\text{.}  \label{7.6}
\end{equation}
\begin{eqnarray}
2\cdot a_0\cdot \rho \text{ }\cdot a &=&-\overline{g}(Krp)+\frac 1e\cdot
[2\cdot a_0\cdot \rho \cdot g(u,a)+g(u,\overline{g}(Krp))]\cdot u-  \notag \\
&&-2\cdot a_0\cdot \rho \text{ }\cdot h^u[(\nabla _ug)(u)]\,\,\,\,\,\text{,}
\label{7.7}
\end{eqnarray}
\begin{equation}
2\cdot a_0\cdot \rho \text{ }\cdot a=-\overline{g}(Krp)-2\cdot a_0\cdot \rho 
\text{ }\cdot h^u[(\nabla _ug)(u)]+f\cdot u\,\,\,\,\,\,\,\text{,}
\label{7.8}
\end{equation}
where 
\begin{equation}
f=\frac 1e\cdot [2\cdot a_0\cdot \rho \cdot g(u,a)+g(u,\overline{g}%
(Krp))]\,\,\,\,\,\,\text{.}  \label{7.9}
\end{equation}

\textit{Special case}: $U_n$- and $V_n$-spaces: $\nabla _\xi g:=0$ for $%
\forall \xi \in T(M)$. 
\begin{equation}
2\cdot a_0\cdot \rho \text{ }\cdot a=-\overline{g}(Krp)+f\cdot
u\,\,\,\,\,\,\,\text{.}  \label{7.10}
\end{equation}

If we apply now the general consideration for the transformation of the
proper time $\tau $ of the vector field $u=d/d\tau $ we can find the
Navier-Stokes equation in a new form by the use of the relations 
\begin{eqnarray*}
a &=&\overline{\alpha }^2\cdot \overline{a}\,\,\,\,\,\,\,\text{,\thinspace
\thinspace \thinspace \thinspace \thinspace \thinspace \thinspace \thinspace
\thinspace \thinspace \thinspace \thinspace }\overline{\alpha }=\overline{%
\alpha }_0\cdot exp[\widetilde{f}\cdot d\tau ]\,\,\,\,\,\,\,\,\,\text{%
,\thinspace \thinspace \thinspace \thinspace \thinspace \thinspace
\thinspace \thinspace \thinspace \thinspace }\overline{\alpha }_0=\text{
const.,} \\
\widetilde{f} &=&\frac 1{2\cdot a_0\cdot \rho }\cdot f\,\,\,\,\text{,}
\end{eqnarray*}
\begin{equation}
a=-\frac 1{2\cdot a_0\cdot \rho }\cdot \overline{g}(Krp)-h^u[(\nabla
_ug)(u)]+\widetilde{f}\cdot u\,\,\,\,\,\,\,\text{,}  \label{7.11}
\end{equation}
\begin{equation*}
a-\widetilde{f}\cdot u=\overline{\alpha }^2\cdot \overline{a}\,\,\,\,\,\text{%
,} 
\end{equation*}
\begin{equation}
\overline{\alpha }^2\cdot \overline{a}=-\frac 1{2\cdot a_0\cdot \rho }\cdot 
\overline{g}(Krp)-h^u[(\nabla _ug)(u)]\,\,\,\,\,\,\text{.}  \label{7.12}
\end{equation}

For the new proper time parameter $\overline{\tau }$ the Navier-Stokes
equation has the form 
\begin{equation}
\overline{a}=-\frac 1{2\cdot a_0\cdot \rho }\cdot \frac 1{\overline{\alpha }%
^2}\cdot \overline{g}(Krp)-\frac 1{\overline{\alpha }^2}\cdot h^u[(\nabla
_ug)(u)]\,\,\,\,\,\,\,\,\text{.}\,\,  \label{7.13}
\end{equation}

Since the metric tensor field $\overline{g}$ is an arbitrary given
contravariant metric, corresponding to the covariant metric tensor $g$, we
can introduce a metric $\widetilde{g}:=\overline{\alpha }^2\cdot g$,
conformal to the metric $g$. Then the corresponding contravariant metric $%
\widetilde{\overline{g}}$ will have the form 
\begin{equation}
\widetilde{\overline{g}}=\frac 1{\overline{\alpha }^2}\cdot \overline{g}%
\,\,\,\,\,\,\text{.}  \label{7.14}
\end{equation}

For the new introduced metric $\widetilde{g}$ the Navier-Stokes equation
could be written in the form 
\begin{equation}
\overline{a}=-\frac 1{2\cdot a_0\cdot \rho }\cdot \widetilde{\overline{g}}%
(Krp)-h^u[\frac 1{\overline{\alpha }^2}\cdot (\nabla _ug)(u)]\,\,\,\,\,\,\,%
\text{.}  \label{7.15}
\end{equation}

On the other side, 
\begin{eqnarray}
u &=&\overline{\alpha }\cdot \overline{u}\,\,\,\,\,\,\,\text{,\thinspace
\thinspace \thinspace \thinspace \thinspace \thinspace \thinspace \thinspace
\thinspace \thinspace }h^u=h^{\overline{u}}\,\,\,\,\,\,\,\text{,}
\label{7.16} \\
\nabla _ug &=&\nabla _{\overline{\alpha }\cdot \overline{u}}g=\overline{%
\alpha }\cdot \nabla _{\overline{u}}g\,\,\,\,\,\text{,\thinspace \thinspace
\thinspace \thinspace \thinspace \thinspace }  \label{7.17} \\
\text{\thinspace }(\nabla _ug)(u) &=&(\nabla _{\overline{\alpha }\cdot 
\overline{u}}g)(\overline{\alpha }\cdot \overline{u})=\overline{\alpha }%
^2\cdot (\nabla _{\overline{u}}g)(\overline{u})\,\,\,\,\,\,\,\,\text{,}
\label{7.18} \\
\frac 1{\overline{\alpha }^2}\cdot (\nabla _ug)(u) &=&(\nabla _{\overline{u}%
}g)(\overline{u})\,\,\,\,\text{.\thinspace \thinspace \thinspace \thinspace
\thinspace }  \label{7.19}
\end{eqnarray}

By the use of the above relations the Navier-Stokes equation takes the form 
\begin{equation}
\overline{a}=-\frac 1{2\cdot a_0\cdot \rho }\cdot \widetilde{\overline{g}}%
(Krp)-h^{\overline{u}}[(\nabla _{\overline{u}}g)(\overline{u}%
)]\,\,\,\,\,\,\,\,\,\,\,\text{,}  \label{7.20}
\end{equation}
which is exactly the generalization of the Euler equation for $(L_n,g)$%
-spaces. In $U_n$- and $V_n$-spaces $(\nabla _{\overline{u}}g=0)$ and for $%
a_0:=\frac 12\cdot \varepsilon $ with \thinspace \thinspace $(\varepsilon
=\mp 1)$ the equation takes its standard form 
\begin{equation}
\overline{a}=\varepsilon \cdot \frac 1\rho \cdot \widetilde{\overline{g}}%
(Krp)\,\,\,\,  \label{7.21}
\end{equation}
or in a co-ordinate basis 
\begin{equation}
\overline{a}^i=\varepsilon \cdot \frac 1\rho \cdot \widetilde{g}\,^{ij}\cdot
p_{,j}\,\,\,\,\,\,\,\,\text{,\thinspace \thinspace \thinspace \thinspace
\thinspace \thinspace \thinspace \thinspace \thinspace \thinspace \thinspace
\thinspace \thinspace }\varepsilon =\mp 1\,\,\,\,\,\,\text{.}  \label{7.22}
\end{equation}

\textit{Remark}. The sign of $\varepsilon $ is depending on the sign of the
pressure $p$ defined by different authors as extrovert (acting out) or
introvert (acting into) quantity with respect to a dynamical system.

\section{Conclusion}

In this paper the Euler equation for a perfect fluid is considered as a
special case of Navier-Stokes' equation in spaces with affine connections
and metrics. By the use of a special type of a Lagrangian density the
Navier-Stokes equations are obtained. It turns out that these equations
could be considered as equations for a perfect fluid in spaces with one
affine connection and a metric [$(L_n,g)$-spaces] and only in special cases
of spaces with contravariant and covariant affine connections and metrics [$(%
\overline{L}_n,g)$-spaces].

The Euler equation is obtained by the use of entirely unconstrained
Lagrangian formalism represented by the method of Lagrangians with covariant
derivatives (MLCD). The additional conditions by this method are related to
the transformation properties of the proper time of a mass element
(particle) moving in space-time. They are not related to any constraints of
the variational principle used for finding out the Euler-Lagrange equations
for a perfect fluid. Therefore, we can consider unconstrained variational
principles in hydrodynamics but under the condition for more careful use of
the notion of proper time (or of time parameter) and its transformation
properties. It seams that the notion of proper time (as the notion of time
in general) could be much more important in physics than assumed until now.

\end{document}